\documentclass[lettersize,journal,romanappendices]{IEEEtran}
\usepackage{amsmath,amsfonts}
\usepackage{algorithmic}
\usepackage{array}
\usepackage[caption=false,font=normalsize,labelfont=sf,textfont=sf]{subfig}
\usepackage{textcomp}
\usepackage{stfloats}
\usepackage{url}
\usepackage{verbatim}
\usepackage{graphicx}
\usepackage{cite}
\usepackage{bm}
\usepackage{diagbox}
\usepackage{tikz}
\usepackage{mathtools}
\usepackage{blkarray, bigstrut}
\usepackage{booktabs}

\def\BibTeX{{\rm B\kern-.05em{\sc i\kern-.025em b}\kern-.08em
    T\kern-.1667em\lower.7ex\hbox{E}\kern-.125emX}}
\usepackage{balance}

\newtheorem{theorem}{Theorem}

\newtheorem{lemma}[theorem]{Lemma}

\newtheorem{remark}[theorem]{Remark}
\newtheorem{example}{Example}

\newtheorem{corollary}[theorem]{Corollary}

\newcommand{\C}{\mathbb C}

\newcommand{\Z}{\mathbb Z}
\newcommand{\F}{\mathbb F}

\newcommand{\ket}[1]{|{#1}\rangle}
\newcommand{\bra}[1]{\langle{#1}|}

\begin{document}
\title{Mitigating Coherent Noise by Balancing\\ Weight-2 $Z$-Stabilizers}
\author{Jingzhen Hu$^\ast$, Qingzhong Liang$^\ast$, \emph{Graduate Student Member, IEEE}, Narayanan Rengaswamy, \emph{Member, IEEE}, and Robert Calderbank, \emph{Fellow, IEEE}
\thanks{$^\ast$The first two authors contributed equally to this work.\\
This work was supported in part by the National Science Foundation (NSF) under Grant CCF-2106213 and Grant CCF-1908730. This article was presented in part at the 2021 IEEE International Symposium on Information Theory \cite{hu2021css}. \\
Jingzhen Hu, Qingzhong Liang, and Robert Calderbank are with the Department of Mathematics, Duke University, Durham, NC 27708, USA (e-mail: jingzhen.hu@duke.edu, qingzhong.liang@duke.edu, robert.calderbank@duke.edu). \\
Narayanan Rengaswamy is with the Department of Electrical and Computer Engineering, University of Arizona, Tucson, AZ 85721, USA (e-mail: narayananr@arizona.edu). Most parts of this work were conducted when Narayanan Rengaswamy was with the Department of Electrical and Computer Engineering, Duke University, Durham, NC 27708, USA.
}}

\markboth{IEEE TRANSACTIONS ON INFORMATION THEORY}%
{Mitigating Coherent Noise by Balancing Weight-2 $Z$-Stabilizers}

\maketitle

\begin{abstract}
	Physical platforms such as trapped ions suffer from coherent noise that does not follow a simple stochastic model. Stochastic errors in quantum systems occur randomly but coherent errors are more damaging since they can accumulate in a particular direction. We consider coherent noise acting transversally, giving rise to an effective error which is a $Z$-rotation on each qubit by some angle $\theta$. 
		Rather than address coherent noise through active error correction, we investigate passive mitigation through decoherence free subspaces. In the language of stabilizer codes, we require the noise to preserve the code space, and to act trivially (as the logical identity operator) on the protected information. Thus, we develop necessary and sufficient conditions for all transversal $Z$-rotations to preserve the code space of a stabilizer code. These conditions require the weight-$2$ $Z$-stabilizers to cover all the qubits that are in the support of the $X$-component of some stabilizer.
		Furthermore, the weight-$2$ $Z$-stabilizers generate a direct product of single-parity-check codes with even block length. By adjusting the sizes of these components, we are able to construct a large family of QECC codes oblivious to coherent noise, one that includes the $[[4L^2, 1, 2L]]$ Shor codes.
		The Shor codes are examples of constant excitation codes, where logical qubits are encoded as a code state that is a sum of physical states indexed by binary vectors with the same weight.  Constant excitation codes are oblivious to coherent noise since a transversal $Z$-rotation acts as a global phase. We prove that a CSS code is oblivious to coherent noise if and only if it is a constant excitation code, and that if the code is error-detecting, then the (constant) weights in different cosets of the $X$-stabilizers are identical.
\end{abstract}


\section{Introduction}
	Quantum error correction is essential to developing scalable and fault-tolerant quantum computers.
	The theory of stabilizer and subsystem codes has led to several promising error correction schemes that provide resilience to quantum noise.
	In quantum systems, noise can broadly be classified into two types -- stochastic and coherent errors.
	Stochastic errors occur randomly and do not accumulate over time along a particular direction. Coherent errors may be viewed as rotations about a particular axis, and can be more damaging, since they can accumulate coherently over time~\cite{Iverson-njp20}.
	As quantum computers move out of the lab and become generally programmable, the research community is paying more attention to coherent errors, and especially to the decay in coherence of the effective induced logical channel~\cite{Beale-prl18,Huang-pra19}.
	It is natural to consider coherent noise acting \emph{transversally}, where the effect of the noise is to implement a separate unitary on each qubit.
	Consider, for example, an $n$-qubit physical system with a uniform background magnetic field acting on the system according to the Hamiltonian $H = Z_{1} +  Z_{2} + \cdots +  Z_{n}$, where $ Z_{i}$ denotes the Pauli $Z$ operator on the $i^{\text{th}}$ qubit.
	Then the effective error is a (unitary) $Z$-rotation on each qubit by some (small) angle $\theta$, i.e., $\exp(\imath \theta H) = \exp(\imath \theta Z)^{\otimes n}$, where $\imath = \sqrt{-1}$.
	
	While it is possible to address coherent noise through active error correction, it can be more economical to passively mitigate such noise through decoherence free subspaces (DFSs)~\cite{Kempe-pra01,alber2001stabilizing}.
	In such schemes, one designs a computational subspace of the full $n$-qubit Hilbert space which is unperturbed by the noise.
	In the language of stabilizer codes, we require the noise to preserve the code space, and to act trivially (as the logical identity operator) on the protected information.
	Inspired by the aforementioned Hamiltonian, which is physically motivated by technologies such as trapped-ion systems, we develop conditions for \emph{all} transversal $Z$-rotations to preserve the code space of a stabilizer code, i.e., $\exp(\imath\theta H) \rho \exp(\imath\theta H)^\dagger = \rho$ for all code states $\rho$ in the stabilizer code. 
	When all angles preserve the code space, the logical action must be trivial for any error-detecting stabilizer code (see Appendix~\ref{sec:proof_logical_identity}).
	The conditions we derive build upon previous work deriving necessary and sufficient conditions for a given transversal $Z$-rotation in the Clifford hierarchy~\cite{gottesman1999demonstrating,Cui-physreva17,Rengaswamy-pra19} to preserve the code space of a stabilizer code~\cite{Opt}.
	The key challenge is handling the trigonometric constraints, and we exploit the celebrated MacWilliams Identities in classical coding theory for this purpose~\cite{Mac}.
	Our main result is a structure theorem that depends on technical arguments which might be of independent interest to classical coding theorists.
	
	The structure theorem forces a product structure on a stabilizer code that is oblivious to coherent noise. Given any even $M$, and any stabilizer code on $t$ qubits, we construct a product code on $Mt$ qubits that is oblivious to coherent noise. The $Mt$ qubits are partitioned into $t$ blocks of $M$ qubits, with each block supporting a DFS. The product code inherits the distance properties of the initial stabilizer code. Thus, the minimal cost of becoming oblivious to coherent noise is scaling the number of qubits by $2$.
	
	The necessary and sufficient conditions for a stabilizer code to be oblivious to coherent noise require the product code structure, resulting in a code rate less than $1/2$. To relax the restrictions, we can consider stabilizer codes that are preserved by all the transversal $Z$-rotations through angle $\pi/2^l$ up to some finite integer $l$, inducing the logical identities. The necessary and sufficient conditions for such error-mitigating codes can be described through the generator coefficient framework \cite{hu2021designing,hu2021climbing} by requiring the generator coefficient corresponding to the trivial syndrome and the trivial $Z$-logical (logical identity) to have norm $1$.

	The paper is organized as follows. Section \ref{sec:main_res} reviews the major technical contributions. Section \ref{sec:prelims} introduces notation and reviews background results. In particular, Section \ref{subsec:gen_enc_map} introduces the general encoding map for CSS codes with arbitrary signs. Section \ref{sec:divisibility} relates divisibility of weights in classical codes to a particular trigonometric identity. Section \ref{sec:transversal_Z_rot} connects stabilizer codes oblivious to coherent noise with a general form of this identity. Section \ref{sec:weight_2_Zs} derives our main result, the structure theorem for stabilizer codes oblivious to coherent noise, Section \ref{sec:construction_method} provides constructions. Section \ref{sec:conclusion} concludes the paper and discusses directions for future work.  
    
    	\section{Discussion of Main Results}
	\label{sec:main_res}
	The introduction of magic state distillation by Bravyi and Kitaev \cite{bravyi2005universal} led to the construction of a sequence of CSS codes~\cite{Calderbank-physreva96,Steane-physreva96}, where the code space is preserved by a transversal $Z$-rotation of the underlying physical space \cite{bravyi2005universal,reichardt2005quantum,anwar2012qutrit,campbell2012magic,bravyi2012magic,landahl2013complex,campbell2017unified,haah2018codes,Haah-pra18,krishna2019towards,Vuillot-arxiv19}. 
	The approach in each paper is to examine the action of a transversal $Z$-rotation on the basis states of a CSS code. This approach results in \emph{sufficient conditions} for a transversal $Z$-rotation to realize a logical operation on the code space. 
	
	In contrast, we derive \emph{necessary and sufficient conditions} by examining the action of the transversal $Z$-rotation on the stabilizer group that determines the code. Thus we study the code space by studying the symmetries of the code space. We start from Rengaswamy \emph{et al.} \cite{Opt} which derived necessary and sufficient conditions for a stabilizer code to be preserved by a transversal $\pi/2^l$ rotation. Note that the condition $l\ge 2$ corresponds to a non-Clifford physical operator. In order to state the result we need to use the notation introduced in Section \ref{sec:prelims}. 
	
	A Hermitian Pauli matrix $\pm E(\bm{a},\bm{b})$ is determined by binary vectors $\bm{a}$ and $\bm{b}$. The $X$-component of $\pm E(\bm{a},\bm{b})$ is $\bm{a}$ and the $Z$-component is $\bm{b}$. A stabilizer group $\mathcal{S}$ is generated by $r$ independent commuting Hermitian Pauli matrices, subject to the requirement that if $E(\bm{a},\bm{b}) \in \mathcal{S}$, then $-E(\bm{a},\bm{b}) \notin \mathcal{S}$. The fixed space $\mathcal{V}(\mathcal{S})$ of $\mathcal{S}$ is an $[[n,n-r]]$ stabilizer code.
	Recall that the Hamming weight $w_H(\bm{v})$ of a binary vector $\bm{v}$ is the number of non-zero entries, and that the support $\mathrm{supp}(\bm{v})$ is the index set of the non-zero entries. Let $\bm{0}$  $(\bm{1})$ be the binary vector with every entry $0$ ($1$). Given $\epsilon E(\bm{a},\bm{b})\in \mathcal{S}$ for some $\epsilon \in \{\pm 1\}$ and $\bm{a}\neq \bm{0}$, define 
	\begin{align}
	    \mathcal{B}(\bm{a}) \coloneqq \{ \bm{z} \in \F_2^{w_H(\bm{a})} : \mathrm{supp}(\bm{z}) \subseteq \mathrm{supp}(\bm{a}) , \epsilon_{\bm{z}} E(\bm{0},\bm{z}) \in \mathcal{S} \}  
	\end{align}
	and 
	\begin{align}
	    \mathcal{O}(\bm{a}) \coloneqq \F_2^{w_H(\bm{a})} \setminus \mathcal{B}(\bm{a}),
	\end{align}
	
	\begin{remark} \label{rem:z_notation}
	\normalfont
	  To simplify notation, we shall sometimes view $\bm{z}$ as a subset of $ \mathrm{supp}(\bm{a})$, sometimes as a subset of the $n$ qubits, and sometimes as a binary vector either of length $w_H(\bm{a})$ or of length $n$ (where entries outside $\mathrm{supp}(\bm{a})$ are set equal to zero). The meaning will be clear from the context.
	\end{remark}
	
	The necessary and sufficient conditions derived by Rengaswamy \emph{et al.} \cite{Opt} are expressed as two trigonometric constraints on weights of pure $Z$-stabilizers in $\mathcal{S}$.
	
		\begin{theorem}[Rengaswamy et al.~\cite{Opt}]
		\label{thm:transversal_Z_rot}
		Transversal $\pi/2^l$  $Z$-rotation ($l \ge 2$) preserves $\mathcal{V}(\mathcal{S})$ if and only if for $\epsilon E(\bm{a},\bm{b}) \in \mathcal{S}$ with $\bm{a} \neq \bm{0}$,
		\begin{align}
		\sum_{\bm{v} \in \mathcal{B}(\bm{a})} \epsilon_{\bm{v}} \left( \imath \tan\frac{2\pi}{2^{l}} \right)^{w_H(\bm{v})} & = \left( \sec\frac{2\pi}{2^{l}} \right)^{w_H(\bm{a})}, \label{theidentity}\\
		\sum_{\bm{v} \in \mathcal{B}(\bm{a})} \epsilon_{\bm{v}} \left( \imath \tan\frac{2\pi}{2^{l}} \right)^{w_H(\bm{v} \oplus \bm{\omega})} & = 0 \quad \text{for\ all}\ \bm{\omega} \in  \mathcal{O}(\bm{a}).
		\label{secondidentity}
		\end{align}      
		Here, $\epsilon_{\bm{v}} \in \{ \pm 1 \}$ is the sign of $E(0,{\bm{v}})$ in the stabilizer group $\mathcal{S}$, and $\oplus$ denotes the binary (modulo $2$) sum of vectors. 
	\end{theorem}
	The theorem reveals that the interaction of transversal physical operators and code states depends very strongly on the signs of pure $Z$-stabilizers. Note that the sign $\epsilon_{\bm{v}}$ of the pure $Z$-stabilizer $\epsilon_{\bm{v}}E(\bm{0,\bm{v}})$ takes the form $\epsilon_{\bm{v}} = (-1)^{\bm{y} \bm{v}^T}$ for $\bm{y}\in \F_2^n$. Note that vectors from the same coset of $\mathcal{C}_1$ (the group of logical $X$ operators) determine the same signs. It is useful to think of $y\in\F_2^n$ as a fixed vector when we extend signs to Pauli matrices outside the stabilizer group.
	
	A stabilizer code is oblivious to coherent noise if and only if transversal $\pi/2^l$ $Z$-rotation preserves the code space $\mathcal{V}(\mathcal{S})$ for all $l\ge 2$ (see Appendix \ref{sec:proof_logical_identity}). We prove that the trigonometric conditions \eqref{theidentity} and \eqref{secondidentity} imply the existence of a large number of weight $2$ $Z$-stabilizers supported on 
	\begin{equation}
	    \Gamma = \bigcup_{\epsilon E(\bm{a},\bm{b}) \in \mathcal{S}} \mathrm{supp}(\bm{a}).
	\end{equation}
	We define a graph with vertex set $\Gamma$, where a vertex corresponds to a qubit of the code and two vertices are joined by an edge if there exists a weight $2$ $Z$-stabilizer involving these two qubits. Let $\Gamma_1,\dots,\Gamma_t$ be the connected components of this graph and let $|\Gamma_k| = N_k$. The weight $2$ $Z$-stabilizers supported on $\Gamma_k$ take the form 
	\begin{equation}
	    (-1)^{\bm{y_k} \bm{v}^T} E(\bm{0},\bm{v}) \text{ where } \bm{y_k} =\bm{y}\big|_{\Gamma_k}.
	\end{equation}
	Here $\bm{y}\big|_{\Gamma_k}$ represents the restriction of $\bm{y}$ to $\Gamma_k$. (In $\bm{y_k} \bm{v}^T$, we add zeros to $\bm{y_k}$ appropriately.)
	Our main result is 
	\begin{theorem} 
	A transversal $\pi/2^l$ $Z$-rotation preserves the stabilizer code for all $l\ge 2$ if and only if for every $\epsilon E(\bm{a},\bm{b}) \in \mathcal{S}$ with $\bm{a}\neq \bm{0}$, 
	\begin{enumerate}
	    \item $\mathrm{supp}(\bm{a})$ is the disjoint union of components $\Gamma_k \subseteq \mathrm{supp}(\bm{a})$,
	    \item $N_k$ is even and $w_H(\bm{y_k})= {N_k}/{2}$ for all $k$ such that $\Gamma_k \subseteq \mathrm{supp}(\bm{a})$.
	\end{enumerate}
\end{theorem}
	
	Note that for every $\epsilon E(\bm{a},\bm{b})\in \mathcal{S}$ we have $\bm{a}\big|_{\Gamma_k} = \bm{0}$ or $\bm{1}$ for $k=1,\dots,t$. Hence Theorem \ref{thm:trans_Z_all_l} forces a product structure on a stabilizer code that is oblivious to coherent noise. It also provides constraints on the signs of weight $2$ $Z$-stabilizers.  
		
	\begin{figure}[h!]
		\centering
		
		\includegraphics[scale=0.45]{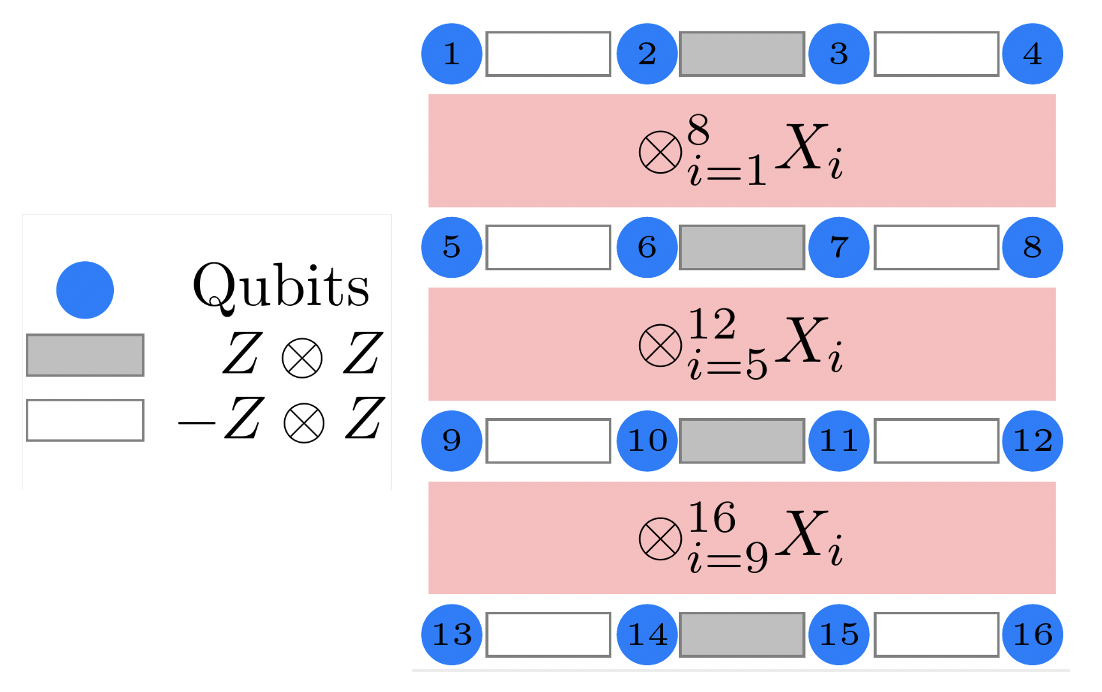}	    
		
		\caption{The $[[16,1,4]]$ Shor code constructed by concatenating the $[[4,1]]$ bit-flip code and the $[[4,1]]$ phase-flip code. The filled circles represent physical qubits, the white (resp. gray filled) squares represent weight-$2$ $Z$-stabilizers with negative (resp. positive) sign, and the three large filled rectangles represent weight-$8$ $X$-stabilizers.} 
		\label{fig:Shor16}
	\end{figure}
	\begin{example}
	\label{exam:Shor}
	\normalfont
	    The $[[16,1,4]]$ Shor code is shown in Fig. \ref{fig:Shor16}, and it follows from Theorem \ref{thm:trans_Z_all_l} that this code is oblivious to coherent noise. The graph on $\Gamma$ has four connected components, and the component $\Gamma_k$ is simply the $k$-th row of the $4\times 4$ array. Condition $(1)$ is satisfied since every $X$ stabilizer is the sum of an even number of rows. Condition $(2)$ is satisfied since the choice $\bm{y_k} = [0,1,1,0]$ for $k=1,2,3,4$ properly accounts for the signs of $Z$-stabilizers. Observe that $[[16,1,4]]$ is also a constant excitation code (defined in Sec.~\ref{subsec:gen_enc_map}). The quotient space $\mathcal{C}_1/\mathcal{C}_2=\{0, \bm{w}=(1 0 0 0)\otimes (1 1 1 1)\}$, where $\mathcal{C}_2$ defines the $X$-stabilizers and $\mathcal{C}_1$ defines the logical $X$ operators. Under the general encoding map, the codewords are
	   \begin{equation}
	       \ket{\overline{0}}=\frac{1}{2\sqrt{2}}\sum_{\bm{x}\in \mathcal{C}_2}\ket{\bm{x}\oplus \bm{y}}  \text{ and } \ket{\overline{1}}=\frac{1}{2\sqrt{2}}\sum_{\bm{x}\in \mathcal{C}_2}\ket{\bm{w}\oplus \bm{x}\oplus \bm{y}}. 
	   \end{equation}
	   
	   The restriction of $\bm{w}$ and $\bm{x}\in \mathcal{C}_2$ to the $k$-th row is either $\bm{0}$ and $\bm{1}$. Since  $w_H(\bm{y_k})=2=\frac{4}{2}$, we have $w_H(\bm{x}\oplus \bm{y})=w_H(\bm{w} \oplus \bm{x}\oplus \bm{y})=8$ for all $\bm{x}\in \mathcal{C}_2$.
	   \end{example}
	   We show that a CSS code is oblivious to coherent noise if and only if it is a constant excitation code (Corollary~\ref{coro:ce_is_nece}). Sufficiency is straightforward since a transversal $Z$-rotation acts as a global phase. 
	   Given a non-degenerate stabilizer code preserved by a diagonal physical gate, we have used the mathematical framework of  generator coefficients to show there is an equivalent CSS code preserved by the same diagonal physical gate and inducing the same logical gate (for more details, see \cite{hu2021designing}).
	   Ouyang \cite{Ouyang-arxiv20b,ouyang2021avoiding} observed that one can construct constant excitation codes by concatenating a stabilizer code with the dual rail code \cite{knill2001scheme}. His original paper was independent of and contemporaneous with our original paper \cite{hu2020mitigating}. After we shared our results he realized that he could connect his dual rail construction to stabilizer code \cite{PC}.
        \section{Preliminaries and Notation}
	\label{sec:prelims}
	
	\subsection{The MacWilliams Identities}
    \label{subsec:prem_Mac_RM}
    Let $\F_2 = \{0,1\}$ denote the binary field. We denote the Hamming weight of a binary vector $\bm{v}$ by $w_H(\bm{v})$. 
    The weight enumerator of a binary linear code $\mathcal{C} \subset \F_2^m$ is the polynomial
    \begin{equation}
    P_{\mathcal{C}}(x,y) = \sum_{\bm{v}\in \mathcal{C}} x^{m-w_H\left(\bm{v}\right)}y^{w_H\left(\bm{v}\right)}.
    \end{equation}
    The MacWilliams Identities \cite{Mac} relate the weight enumerator of a code $\mathcal{C}$ to that of the dual code $\mathcal{C}^\perp$, and are given by
    \begin{equation}
    P_{\mathcal{C}}(x,y) = \frac{1}{|\mathcal{C}^\perp|} P_{\mathcal{C}^\perp}(x+y,x-y).
    \end{equation}
    We frequently make the substitution $x=\cos\frac{2\pi}{2^l}$ and $y=-\imath\sin\frac{2\pi}{2^l}$, and we define
    \begin{align}
    P[\mathcal{C}] 
    &\coloneqq P_{\mathcal{C}}\left(\cos\frac{2\pi}{2^l},-\imath\sin\frac{2\pi}{2^l}\right) \\
    &= \sum_{\bm{v}\in \mathcal{C}} \left(\cos\frac{2\pi}{2^l}\right)^{m-w_H(\bm{v})}\left(-\imath\sin\frac{2\pi}{2^l}\right)^{w_H(\bm{v})}. \label{eqn:key_sub_Mac}
    \end{align}

    \subsection{The Pauli Group}
    \label{subsec:prem_paulis}
    Let $N=2^n$. 
    Any $2\times 2$ Hermitian matrix can be uniquely expressed as a real linear combination of the four single qubit Pauli matrices/operators
    \begin{equation}
    I_2 \coloneqq \begin{bmatrix}
    1 & 0\\
    0 & 1
    \end{bmatrix},~  
    X \coloneqq \begin{bmatrix}
    0 & 1\\
    1 & 0
    \end{bmatrix},~  
    Z \coloneqq  \begin{bmatrix}
    1 & 0\\
    0 & -1
    \end{bmatrix}, ~ 
    Y=\imath XZ,
    \end{equation}
    where $\imath=\sqrt{-1}$. 
    The operators satisfy 
    $
    X^2= Y^2= Z^2=I_2,~  X Y=- Y X,~  X Z=- Z X,~ \text{ and }  Y Z=- Z Y.
    $
    
    Let $A \otimes B$ denote the Kronecker product (tensor product) of two matrices $A$ and $B$. Given vectors $\bm{a}=[a_1,a_2,\dots,a_n]$ and $\bm{b}=[b_1,b_2,\dots,b_n]$ with $a_i,b_j =0$ or $1$, we define the operators
    \begin{align}
    D(\bm{a},\bm{b})&\coloneqq X^{a_1} Z^{b_1}\otimes  X^{a_2} Z^{b_2}\otimes \cdots \otimes  X^{a_n} Z^{b_n},\\
    E(\bm{a},\bm{b}) &\coloneqq\imath^{\bm{a}\bm{b}^T \pmod 4}D(\bm{a},\bm{b}).
    \end{align}
    
    We often abuse notation and write $\bm{a}, \bm{b} \in \F_2^n$, though entries of vectors are sometimes interpreted in $\mathbb{Z}_4 = \{ 0,1,2,3 \}$. Note that $D(\bm{a},\bm{b})$ can have order $1,2$ or $4$ (order means the smallest positive integer $h$ such that $D(\bm{a},\bm{b})^{h}=I_N$), but $E(\bm{a},\bm{b})^2=\imath^{2\bm{a}\bm{b}^T}D(\bm{a},\bm{b})^2=\imath^{2ab^T}( \imath^{2\bm{a}\bm{b}^T} I_N)=I_N$. The $n$-qubit \textit{Pauli group} is defined as
    \begin{equation}
    \mathcal{P}_n \coloneqq\{\imath^\kappa D(\bm{a},\bm{b}): \bm{a},\bm{b}\in \F_2^n, \kappa=0,1,2,3 \}.
    \end{equation}
    The $n$-qubit Pauli matrices form an orthonormal basis for the vector space of $N\times N$ complex matrices $\C^{N\times N}$ under the normalized Hilbert-Schmidt inner product $\langle A,B\rangle \coloneqq \mathrm{Tr}(A^\dagger B)/N$.
    
    We will use the \textit{Dirac notation}, $|\cdot \rangle$ to represent the basis states of a single qubit in $\C^2$. For any $\bm{v}=[v_1,v_2,\cdots, v_n]\in \F_2^n$, we define $|\bm{v}\rangle=|v_1\rangle\otimes|v_2\rangle\otimes\cdots\otimes|v_n\rangle$, the standard basis vector in $\C^N$ with $1$ in the position indexed by $\bm{v}$ and $0$ elsewhere. 
    We write the Hermitian transpose of $|\bm{v}\rangle$ as $\langle \bm{v}|=|\bm{v}\rangle^\dagger$. 
    We may write an arbitrary $n$-qubit quantum state as $|\psi\rangle=\sum_{\bm{v}\in \F_2^n} \alpha_{\bm{v}} |\bm{v}\rangle \in \C^N$, where $\alpha_{\bm{v}}\in \C$ and $\sum_{\bm{v}\in\F_2^n}|\alpha_{\bm{v}}|^2=1$. The Pauli matrices act on a single qubit as
    \begin{equation}
    X|0\rangle=|1\rangle,  X|1\rangle=|0\rangle,  Z|0\rangle=|0\rangle, \text{ and }  Z|1\rangle=-|1\rangle.
    \end{equation}
    
    The symplectic inner product is $\langle [\bm{a},\bm{b}],[\bm{c},\bm{d}]\rangle_S=\bm{a}\bm{d}^T+\bm{b}\bm{c}^T \pmod 2$. Since $ X Z=- Z X$, we have 
    \begin{equation}
    E(\bm{a},\bm{b})E(\bm{c},\bm{d})=(-1)^{\langle [\bm{a},\bm{b}],[\bm{c},\bm{d}]\rangle_S}E(\bm{c},\bm{d})E(\bm{a},\bm{b}).
    \end{equation}

    \subsection{The Clifford Hierarchy}
    \label{subsec:prem_Cliff}
    
    The \textit{Clifford hierarchy} of unitary operators was introduced in \cite{gottesman1999demonstrating}. The first level of the hierarchy is defined to be the Pauli group $\mathcal{C}^{(1)}=\mathcal{P}_n$. For $l\ge 2$, the levels $l$ are defined recursively as 
    \begin{equation}\label{eqn:def_Cliff_hierarchy}
    \mathcal{C}^{(l)}:=\{U\in \mathbb{U}_N: UE(\bm{a},\bm{b})U^\dagger\in \mathcal{C}^{(l-1)}, ~\text{for all}~ E(\bm{a},\bm{b})\in \mathcal{P}_n \},
    \end{equation}
    where $\mathbb{U}_N$ is the group of $N\times N$ unitary matrices. The second level is the Clifford Group~\cite{Gottesman-icgtmp98}, $\mathcal{C}^{(2)}$, which can be generated using the unitaries \textit{Hadamard}, \textit{Phase}, and either of \textit{Controlled-NOT} (C$X$) or \textit{Controlled-$Z$} (C$Z$) defined respectively as
    \begin{equation}
    H\coloneqq\frac{1}{\sqrt{2}}
    \begin{bmatrix}
    1 & 1\\
    1 & -1
    \end{bmatrix},~
    P\coloneqq\begin{bmatrix}
    1 & 0\\
    0 & \imath
    \end{bmatrix},
    \end{equation}
    \begin{align}
    \text{C}Z_{ab}  &\coloneqq \ket{0}\bra{0}_a \otimes (I_2)_b + \ket{1}\bra{1}_a \otimes Z_b,\\
    \text{C}X_{a \rightarrow b}  &\coloneqq \ket{0}\bra{0}_a \otimes (I_2)_b + \ket{1}\bra{1}_a \otimes X_b.
    \end{align}
    
    It is well-known that Clifford unitaries in combination with \emph{any} unitary from a higher level can be used to approximate any unitary operator arbitrarily well~\cite{boykin1999universal}. 
    Hence, they form a universal set for quantum computation. A widely used choice for the non-Clifford unitary is the $T$ gate defined by 
    \begin{equation}
    T:=\begin{bmatrix}
    1 & 0\\
    0 & e^{\frac{i\pi}{4}}
    \end{bmatrix}
    =\sqrt{P}= Z^{\frac{1}{4}}\equiv \begin{bmatrix}
    e^{-\frac{\imath\pi}{8}} & 0\\
    0 & e^{\frac{\imath\pi}{8}}
    \end{bmatrix} 
    =e^{-\frac{\imath\pi}{8} Z}.
    \end{equation}
    
    \subsection{Stabilizer Codes}
    \label{subsec:prem_stab}
    We define a stabilizer group $\mathcal{S}$ to be a commutative subgroup of the Pauli group $\mathcal{P}_n$, where every group element is Hermitian and no group element is $-I_N$. We say $\mathcal{S}$ has dimension $r$ if it can be generated by $r$ independent elements as $\mathcal{S}=\langle \nu_i E(\bm{c_i},\bm{d_i}): i=1,2,\dots, r \rangle$, where $\nu_i\in\{\pm1\}$ and $\bm{c_i},\bm{d_i}\in \F_2^n$. Since $\mathcal{S}$ is commutative, we must have $\langle [\bm{c_i},\bm{d_i}],[\bm{c_j},\bm{d_j}]\rangle_S=\bm{c_i}\bm{d_j}^T+\bm{d_i}\bm{c_j}^T=0\pmod 2$.
    
    Given a stabilizer group $\mathcal{S}$, the corresponding \textit{stabilizer code} is the fixed subspace $\mathcal{V}(\mathcal{S)}:=\{|\psi\rangle \in \C^N: g|\psi\rangle=|\psi\rangle \text{ for all } g\in \mathcal{S} \}$. 
    We refer to the subspace $\mathcal{V}(\mathcal{S})$ as an $\left[\left[n,k,d\right]\right]$ stabilizer code because it encodes $k:=n-r$ \text{logical} qubits into $n$ \textit{physical} qubits. The minimum distance $d$ is defined to be the minimum weight of any operator in $\mathcal{N}_{\mathcal{P}_n}\left(\mathcal{S}\right)\setminus \mathcal{S}$. Here, the weight of a Pauli operator is the number of qubits on which it acts non-trivially (i.e., as $ X,~Y$ or $ Z$), and $\mathcal{N}_{\mathcal{P}_n}\left(\mathcal{S}\right)$ denotes the normalizer of $\mathcal{S}$ in $\mathcal{P}_n$ defined by 
    \begin{align}
        \mathcal{N}_{\mathcal{P}_n}\left(\mathcal{S}\right) \coloneqq &\{\imath^\kappa E\left(\bm{a},\bm{b}\right)\in \mathcal{P}_n: E\left(\bm{a},\bm{b}\right)E\left(\bm{c},\bm{d}\right)E\left(\bm{a},\bm{b}\right)=
        \nonumber \\
        & E\left(\bm{c}',\bm{d}'\right) \in \mathcal{S} \text{ for all } \nu E\left(\bm{c},\bm{d}\right)\in \mathcal{S}, \kappa\in \Z_4 \} \nonumber\\
       =& \{\imath^\kappa E\left(\bm{a},\bm{b}\right)\in \mathcal{P}_n: E\left(\bm{a},\bm{b}\right)E\left(\bm{c},\bm{d}\right)E\left(\bm{a},\bm{b}\right)= 
       \nonumber \\
       &E\left(\bm{c},\bm{d}\right) \text{ for all } \nu E\left(\bm{c},\bm{d}\right)\in \mathcal{S}, \kappa\in \Z_4 \}.
    \end{align}
    Note that the second equality defines the centralizer of $\mathcal{S}$ in $\mathcal{P}_n$, and it follows from the first since Pauli matrices commute or anti-commute.
    
    For any Hermitian Pauli matrix $E\left(\bm{c},\bm{d}\right)$ and $\nu\in\{\pm 1\}$, the projector $\frac{I_N+\nu E\left(\bm{c},\bm{d}\right)}{2}$ projects on to the $\nu$-eigenspace of $E\left(\bm{c},\bm{d}\right)$. Thus, the projector on to the codespace $\mathcal{V}(\mathcal{S})$ of the stabilizer code defined by $\mathcal{S}=\langle \nu_i E\left(\bm{c_i},\bm{d_i}\right): i=1,2,\dots, r \rangle$ is 
    \begin{equation}
    \Pi_{\mathcal{S}}=\prod_{i=1}^{r}\frac{\left(I_N+\nu_iE\left(\bm{c_i},\bm{d_i}\right)\right)}{2}=\frac{1}{2^r}\sum_{j=1}^{2^r}\epsilon_jE\left(\bm{a_j},\bm{b_j}\right),
    \end{equation}
    where $\epsilon_j\in \{\pm 1 \}$ is a character of the group $\mathcal{S}$, and is determined by the signs of the generators that produce $E(\bm{a_j},\bm{b_j})$: $\epsilon_jE\left(\bm{a_j},\bm{b_j}\right)=\prod_{t\in J\subset \{1,2,\dots,r\} } \nu_t E\left(\bm{c_t},\bm{d_t}\right)$ for a unique $J$.
    
	
    \subsection{CSS Codes}
    \label{subsec:prem_CSS}
    A \textit{CSS (Calderbank-Shor-Steane) code} is a type of stabilizer code with generators that can be separated into strictly $X$-type and $Z$-type operators~\cite{Calderbank-physreva96,Steane-physreva96}. Consider two classical binary codes $\mathcal{C}_1,\mathcal{C}_2$ such that $\mathcal{C}_2\subset \mathcal{C}_1$, and let $\mathcal{C}_1^\perp$, $\mathcal{C}_2^\perp$ denote the dual codes. Note that $\mathcal{C}_1^\perp\subset \mathcal{C}_2^\perp$. Suppose that $\mathcal{C}_2 = \langle \bm{{c}_1},\bm{{c}_2},\dots,\bm{c_{k_2}} \rangle$ is an $[n,k_2]$ code and $\mathcal{C}_1^\perp =\langle \bm{d_1},\bm{d_2}, \dots,\bm{d_{n-k_1}}\rangle$ is an $[n,n-k_1]$ code. Then, the corresponding CSS code has the stabilizer group 
	\begin{align}
	\mathcal{S} 
	&=\langle \nu_{(\bm{c_i},\bm{0})} E\left(\bm{c_i},\bm{0}\right), \nu_{(\bm{0},\bm{d_j})} E\left(\bm{0},\bm{d_j}\right)\rangle_{\substack{i\in\{1,\dots,k_2\}, \\ j\in\{1,\dots,n-k_1\}}} \nonumber\\
	&=\{\epsilon_{(\bm{a},\bm{0})} \epsilon_{(\bm{0},
	\bm{b})} E\left(\bm{a},\bm{0}\right)E\left(\bm{0},\bm{b}\right): \bm{a}\in \mathcal{C}_2, \bm{b}\in \mathcal{C}_1^\perp\}, 
	\end{align}
	where $\nu_{(\bm{c_i},\bm{0})},\nu_{(\bm{d_j},\bm{0})},\epsilon_{(\bm{a},\bm{0})},\epsilon_{(\bm{0},
	\bm{b})}  \in\{\pm 1 \}$.
    The CSS code projector can be written as the product:
	\begin{equation}
	\Pi_{\mathcal{S}} 
	=\Pi_{\mathcal{S}_X}\Pi_{\mathcal{S}_Z},
	\end{equation}
	where 
	\begin{equation}
	\Pi_{\mathcal{S}_X} \eqqcolon \prod_{i=1}^{k_2} \frac{(I_N+\nu_{(\bm{c_i},\bm{0})} E(\bm{c_i},\bm{0}))}{2} = \frac{\sum_{\bm{a}\in \mathcal{C}_2} \epsilon_{(\bm{a},\bm{0})} E(\bm{a},\bm{0})}{|\mathcal{C}_2|},
	\end{equation}
	and 
	\begin{equation}
	\Pi_{\mathcal{S}_Z} \eqqcolon \prod_{j=1}^{n-k_1} \frac{(I_N + \nu_{(\bm{0},\bm{d_j})} E(\bm{0},\bm{d_j}))}{2} = \frac{\sum_{\bm{b}\in \mathcal{C}_1^\perp} \epsilon_{(\bm{0},
	\bm{b})}  E(\bm{0},\bm{b})}{|\mathcal{C}_1^\perp|}.
	\end{equation}
	
	If $\mathcal{C}_1$ and $\mathcal{C}_2^\perp$ can correct up to $t$ errors, then $S$ defines an $\left[\left[n, k, d\right]\right]$ CSS code, $k = k_1-k_2$, with $d\ge 2t+1$, which we will represent as CSS($X,\mathcal{C}_2;Z,\mathcal{C}_1^\perp$). If $G_2$ and $G_1^\perp$ are the generator matrices for $\mathcal{C}_2$ and $\mathcal{C}_1^\perp$ respectively, then the $(n-k_1+k_2)\times (2n)$ matrix
	\begin{equation}
	G_{\mathcal{S}}=\left[\begin{array}{c|c}
	G_2 &  \\ \hline
	&  G_1^\perp
	\end{array} \right]
	\end{equation}
	generates $\mathcal{S}$. The codespace defined by the stabilizer group $\mathcal{S}$ is $ \mathcal{V}(\mathcal{S}):=\{\ket{\psi}\in \C^N : g\ket{\psi} = \ket{\psi} \text{ for all } g\in \mathcal{S} \}$.

    \subsection{Encoding Map for CSS codes}
    \label{subsec:gen_enc_map}
    Given an $[[n,k,d]]$ CSS($X,\mathcal{C}_2;Z,\mathcal{C}_1^\perp$) code with all positive signs, let $G_{\mathcal{C}_1/\mathcal{C}_2} \in \mathbb{F}_2^{k \times n}$ be a matrix that generates for all coset representatives for $\mathcal{C}_2$ in $\mathcal{C}_1$ (note that the choice of coset representatives is not unique). The canonical encoding map $f:\mathbb{F}_2^k \to \mathcal{V}(\mathcal{S})$ is given by 
	$
	\ket{\overline{\bm{v}}}\coloneqq f(\ket{\bm{v}}_L) 
	\coloneqq \frac{1}{\sqrt{|\mathcal{C}_2|}} \sum_{\bm{x}\in \mathcal{C}_2}\ket{\bm{v}  G_{\mathcal{C}_1/\mathcal{C}_2} \oplus \bm{x}}$. 
	Changing the signs of stabilizers changes the fixed subspace. 
	Hence we need to modify the encoding map to account for nontrivial signs. Define subspaces $\mathcal{B}$ and $\mathcal{D}$ as below.
	\begin{center}
		\begin{tikzpicture}
		\node (B) at (1,1) {$\mathcal{B}=\{\bm{z}\in \mathcal{C}_1^\perp|\epsilon_{\bm{z}}=1\}$};
		\node (C) at (1,2) {$\mathcal{C}_1^\perp$};
		\path[draw] (B) -- (C);
		\node (Cp) at (3,1) {$\mathcal{C}_1$};
		\node (Bp) at (3,2) {$\mathcal{B}^{\perp}$};
		\path[draw] (Cp) -- (Bp);
		
		\node (D) at (5.5,1) {$\mathcal{D}=\{\bm{x}\in \mathcal{C}_2|\epsilon_{\bm{x}}=1\}$};
		\node (C2) at (5.5,2) {$\mathcal{C}_2$};
		\path[draw] (D) -- (C2);
		\node (C2p) at (7.5,1) {$\mathcal{C}_2^\perp$};
		\node (Dp) at (7.5,2) {$\mathcal{D}^{\perp}$};
		\path[draw] (C2p) -- (Dp);
		
		\end{tikzpicture}
	\end{center}
	We capture sign information through character vectors 
	$\bm{y},~\bm{u} \in \F_2^n$
	(note that the choice of $\bm{y}, \bm{u}$ is unique only up to elements in $\mathcal{C}_1, \mathcal{C}_2^\perp$ respectively) satisfying 
	\begin{equation}
	\mathcal{B}=\mathcal{C}_1^\perp\cap \bm{y}^\perp,\text{or equivalently, } \mathcal{B}^\perp = \langle \mathcal{C}_1,\bm{y}\rangle,
	\end{equation}
	and 
	\begin{equation}
	\mathcal{D} =\mathcal{C}_2 \cap \bm{u}^\perp,\text{or equivalently, }D^\perp = \langle \mathcal{C}_2^\perp,\bm{u}\rangle.
	\end{equation}
    Then, for $ \epsilon_{(\bm{a},\bm{0})} \epsilon_{(\bm{0},\bm{b})} E\left(\bm{a},\bm{0}\right)E\left(\bm{0},\bm{b}\right) \in S$, we have $ \epsilon_{(\bm{a},\bm{0})} = (-1)^{\bm{a}\bm{u}^T}$ and $ \epsilon_{(\bm{0},\bm{b})} = (-1)^{\bm{b}\bm{y}^T}$. 
	
	The canonical bijective map $f:\mathbb{F}_2^k \to \mathcal{V}(\mathcal{S})$ becomes \cite{hu2021designing}
	\begin{equation} \label{eqn:gen_encode_map}
		\ket{\overline{\bm{v}}}= f(\ket{\bm{v}}_L) 
		\coloneqq \frac{1}{\sqrt{|\mathcal{C}_2|}} \sum_{\bm{x}\in \mathcal{C}_2} (-1)^{\bm{x}\bm{u}^T}\ket{\bm{v}  G_{\mathcal{C}_1/\mathcal{C}_2} \oplus \bm{x} \oplus \bm{y}}.
	\end{equation}
	 The CSS code is said to be a \emph{constant excitation code} \cite{zanardi1997noiseless} if, for each fixed $\bm{v}\in\F_2^k$, the weight  $w_H(\bm{v}G_{\mathcal{C}_1/\mathcal{C}_2}\oplus \bm{x} \oplus \bm{y})$ is constant for all $\bm{x}\in \mathcal{C}_2$. Recall that a common kind of coherent noise is modeled by $U=\exp(\imath \theta Z)^{\otimes n}$ for arbitrary $\theta$. When $U$ acts on a $\ket{0}\&\ket{1}$ computational basis state in a constant excitation code, each term in \eqref{eqn:gen_encode_map} generates the same phase term $\exp(\imath \theta w_H(\bm{v}G_{\mathcal{C}_1/\mathcal{C}_2}\oplus \bm{x} \oplus \bm{y}))$, leading to a global phase, which leaves the state invariant. Hence, a constant excitation code is oblivious to coherent noise.

	\section{Divisibility of Weights in Binary Codes} \label{sec:divisibility}

	The defining property of a divisible linear code \cite{Ward} is that codeword weights share a common divisor larger than one. Codes obtained by repeating each coordinate in a shorter code the same number of times are automatically divisible, and they are essentially the only ones for divisors prime to the field size. Examples that are more interesting occur when the divisor is a power of the characteristic. For example, the theorem of Ax \cite{Ax} governing the existence of zeros of polynomials in several variables characterizes divisibility of weights in Reed-Muller codes \cite{Ax,McEliece,macwilliams1977theory,borissov2013mceliece}. 
	
	Divisible codes (in particular Reed-Muller codes) appear in protocols designed for magic state distillation \cite{bravyi2005universal,anwar2012qutrit,campbell2012magic,bravyi2012magic} which achieves universal quantum computation through transversal implementation of Clifford gates and ancillary magic states.  Divisibility tests \cite{landahl2013complex,Vuillot-arxiv19} are introduced to ensure that a quantum error correcting code is preserved by a transversal $\pi/2^l$ $Z$-rotation. We argue in the reverse direction, showing that divisibility of weights is forced by the requirement that the quantum error correcting code is fixed by a transversal gate. We will make repeated use of the following trigonometric identity that is equivalent to code divisibility and may be of independent interest to classical coding theorists.
	
	
	\begin{lemma}\label{all_one_iff}
		Let $\mathcal{C}$ be a binary linear code with block length $m$, where all weights are even. Let $l \ge 2$. Then, 
		\begin{equation}\label{all_one_iff_assump}
		\sum_{\bm{v}\in \mathcal{C}} \left(\imath\tan{\frac{2\pi}{2^l}}\right)^{w_H(\bm{v})}=\left(\sec{\frac{2\pi}{2^l}}\right)^{m}
		\end{equation}
		if and only if $(m-2w_H(\bm{w}))$ is divisible by $2^l$ for all $\bm{w}\in \mathcal{C}^\perp$.
	\end{lemma}
	\begin{IEEEproof}
		We rewrite (\ref{all_one_iff_assump}) as
		\begin{equation}\label{all_one_iff_before_Mac}
		P[\mathcal{C}] = \sum_{\bm{v}\in \mathcal{C}} \left(\cos\frac{2\pi}{2^l}\right)^{m-w_H(\bm{v})}\left(\imath\sin{\frac{2\pi}{2^l}}\right)^{w_H(\bm{v})} = 1.
		\end{equation}
		Let $t_{+}\coloneqq \cos\frac{2\pi}{2^l}+\imath\sin{\frac{2\pi}{2^l}}$ and $t_{-}\coloneqq \cos\frac{2\pi}{2^l}-\imath\sin{\frac{2\pi}{2^l}}$. After applying the MacWilliams identities, (\ref{all_one_iff_before_Mac}) becomes
		\begin{equation} \label{all_one_iff_middle_step}
		\frac{1}{|\mathcal{C}^\perp|}P_{\mathcal{C}^\perp}\left(t_{+},t_{-}\right) = 1.
		\end{equation}
		Since $\left(\cos\theta + \imath\sin\theta\right)\left(\cos\theta - \imath\sin\theta\right) = 1 \text{ for all } \theta$, we may rewrite (\ref{all_one_iff_middle_step}) as
		\begin{equation}
		\frac{1}{|\mathcal{C}^\perp|}\sum_{\bm{w}\in \mathcal{C}^\perp} t_{+}^{m-w_H(\bm{w})} t_{-}^{w_H(\bm{w})} = 1,
		\end{equation}
		which may be further simplified to 
		\begin{equation}\label{all_one_iff_middle_2}
		\frac{1}{|\mathcal{C}^\perp|}\sum_{\bm{w}\in \mathcal{C}^\perp} t_{+}^{m-2w_H(\bm{w})} = 1.
		\end{equation}
		Since $\bm{1} \in \mathcal{C}^\perp$, the complement of a codeword in $\mathcal{C}^\perp$ is again a codeword in $\mathcal{C}^\perp$, so we may rewrite (\ref{all_one_iff_middle_2}) as
		\begin{equation}\label{all_one_iff_after_MW}
		\frac{1}{|\mathcal{C}^\perp|}\left[\sum_{\bm{w}\in \mathcal{C}^\perp} t_{+}^{m-2w_H(\bm{w})} + \sum_{\bm{w}\in \mathcal{C}^\perp} t_{+}^{-(m-2w_H(\bm{w}))}\right]= 2.
		\end{equation}
		Since $(\cos\theta + \imath\sin\theta)^n = e^{\imath n\theta}$, for all $\theta$, equation (\ref{all_one_iff_after_MW}) reduces to,
		\begin{equation}\label{last_step}
		\frac{1}{|\mathcal{C}^\perp|}\sum_{\bm{w}\in \mathcal{C}^\perp}\cos\left(\frac{2\left(m-2w_H\left(\bm{w}\right)\right)\pi}{2^l}\right) = 1. 
		\end{equation}
		We observe that equation (\ref{last_step}) is satisfied if and only if each term contributes $1$ to the sum, and this is equivalent to $2^l$ dividing $m-2w_H(\bm{w})$ for all codewords $\bm{w}$ in $\mathcal{C}^\perp$.
	\end{IEEEproof}

	Setting $\mathcal{C} = \mathcal{B}(\bm{a})$ in the above lemma provides insights into the conditions of Theorem~\ref{thm:transversal_Z_rot}.
	
	\section{Transversal $Z$-Rotations}
	\label{sec:transversal_Z_rot}

	Given two binary vectors $\bm{x},\bm{y}$, we write $\bm{x}\preceq \bm{y}$ to mean that the \emph{support} of $\bm{x}$ is contained in the support of $\bm{y}$. We define $\bm{y}|_{\mathrm{supp}(\bm{x})} \in \F_2^{w_H(\bm{x})}$ to be the restriction of $\bm{y}$ to  $\mathrm{supp}(\bm{x})$. Consider the $[[n,n-r]]$ stabilizer code $\mathcal{V}(\mathcal{S})$ determined by the stabilizer group 
	$S=\langle \nu_i E(\bm{c_i},\bm{d_i}) : \nu_i \in\{ \pm 1\},i=1,\cdots,r\rangle$. 
	Recall that given a stabilizer $\epsilon E(\bm{a},\bm{b})$ with $\bm{a} \neq \bm{0}$, we define 
	\begin{align}
	    \mathcal{B}(\bm{a}) = \{\bm{z}\big|_{\mathrm{supp}(\bm{a})} \in \F_2^{w_H(\bm{a})}: \epsilon_{\bm{z}} E\left(\bm{0},\bm{z}\right) \in \mathcal{S} \text{ and } \bm{z} \preceq \bm{a}\}
	\end{align}
	and
	\begin{align}
	 \mathcal{O}(\bm{a}) = \F_2^{w_H(\bm{a})} \setminus \mathcal{B}(\bm{a}) = \{\bm{\omega}\in\F_2^{w_H(\bm{a})}: \bm{\omega}\notin \mathcal{B}(\bm{a})\}.
	\end{align}
	Since $\mathcal{S}$ is commutative, $\bm{1} \in \mathcal{B}(\bm{a})^\perp$, and it follows that all weights in $\mathcal{B}(\bm{a})$ are even. 

    \begin{example}\normalfont
        Consider the $[[16,1,4]]$ Shor code shown in Figure \ref{fig:Shor16}. Setting $E(\bm{a},\bm{0}) = \otimes_{i=1}^8 X_i$, where $X_i$ means Pauli $X$ on the $i$-th qubit, we have  $\mathcal{B}(\bm{a}) = \F_2^2 \otimes \langle [1,1,0,0], [0,1,1,0],[0,0,1,1]\rangle$.  
    \end{example}

	We now consider Theorem \ref{thm:transversal_Z_rot} in the special case $l=2$ (Transversal $T$). Let 
		\begin{equation}
		s = \sum_{\bm{v}\in \mathcal{B}(\bm{a})} \epsilon_{\bm{v}} \imath^{w_H(\bm{v})}.
		\end{equation}
		Since $\tan\frac{\pi}{4}=1$ and $\sec\frac{\pi}{4}=\sqrt{2}$, we may rewrite \eqref{theidentity} as 
		\begin{align}
		s^2 = 2^{w_H(\bm{a})} 
		& = \sum_{\bm{v},\bm{w} \in \mathcal{B}(\bm{a})}\epsilon_{\bm{v}} \epsilon_{\bm{w}} \imath^{w_H(\bm{v})+w_H(\bm{w})}\\
		& = \sum_{\bm{v},\bm{w} \in \mathcal{B}(\bm{a})}\epsilon_{\bm{v}\oplus \bm{w}}  \imath^{w_H(\bm{v} \oplus \bm{w})+2\bm{v} \bm{w}^T}.
		\end{align}
		Changing variables to $\bm{z}=\bm{v}\oplus \bm{w}$ and $\bm{v}$, we obtain
%
        \begin{align}
		2^{w_H(\bm{a})} 
		& = \sum_{\bm{z},\bm{v}\in \mathcal{B}(\bm{a})}\epsilon_{\bm{z}} \imath^{w_H(\bm{z})} \left(-1\right)^{(\bm{z}\oplus \bm{v})\bm{v}^T}\\
		& = \sum_{\bm{z}\in \mathcal{B}(\bm{a})} \epsilon_{\bm{z}} \imath^{w_H(\bm{z})}\sum_{\bm{v}\in \mathcal{B}(\bm{a})} \left(-1\right)^{\bm{z} \bm{v}^T} \\
		& = |\mathcal{B}(\bm{a})|\sum_{\bm{z}\in \mathcal{B}(\bm{a}) \cap \mathcal{B}(\bm{a})^\perp} \epsilon_{\bm{z}} \imath^{w_H(\bm{z})},
		\end{align}
		where the second step follows from $\ \bm{v} \bm{v}^T$ is even. 
		Since $2^{w_H(\bm{a})} = |\mathcal{B}(\bm{a})|\cdot|\mathcal{B}(\bm{a})^\perp|$ and $|\mathcal{B}(\bm{a}) \cap \mathcal{B}(\bm{a})^\perp| \le |\mathcal{B}(\bm{a})^\perp|$, $\mathcal{B}(\bm{a})^\perp$ is contained in $\mathcal{B}(\bm{a})$ and so $\bm{1} \in \mathcal{B}(\bm{a})$. 
		Since $\mathcal{B}(\bm{a})^\perp \subseteq \mathcal{B}(\bm{a})$, it now follows that $\mathcal{B}(\bm{a})$ contains a self-dual code. Since 
		\begin{equation}
		|\mathcal{B}(\bm{a})^\perp| = \sum_{\bm{z}\in \mathcal{B}(\bm{a})^\perp}\epsilon_{\bm{z}} \imath^{w_H(\bm{z})},
		\end{equation}
		we must have $\epsilon_{\bm{z}}=\imath^{w_H(\bm{z})}$
    for all $\bm{z} \in \mathcal{B}(\bm{a})^\perp$.
		
    \begin{remark}\normalfont
    \label{rem:arrive Transversal T thm}
        The above derivation provides the three necessary conditions given in \cite[Theorem 2]{Opt} that are necessary for a stabilizer code to be preserved by the transversal $T$ gate.
        \begin{enumerate}
			\item For each $\epsilon E(\bm{a},\bm{b}) \in \mathcal{S}$ with $\bm{a} \neq \bm{0}$, the Hamming weight $w_H(\bm{a})$ is even.
			\item For each $\epsilon E(\bm{a},\bm{b}) \in \mathcal{S}$ with $\bm{a} \neq \bm{0}$, the binary code $\mathcal{B}(\bm{a})$ contains an $\left[n=w_H(\bm{a}),k=\frac{w_H\left(\bm{a}\right)}{2}\right]$ self-dual code.
			\item For each $\bm{z} \in \mathcal{B}(\bm{a})^\perp$, the sign of the corresponding stabilizer $E(\bm{0},{\bm{z}}) \in \mathcal{S}$ is given by $\imath^{w_H({\bm{z}})}$.
		\end{enumerate}
    \end{remark}
    
	\begin{example}\normalfont \label{[[16,4,2]]-QRM}
		Consider the $[[16,4,2]]$ code that is a member of the $[[2^m,\binom{m}{1}, 2]]$ quantum Reed-Muller (QRM) family constructed in~\cite{Opt}. 
		It is the CSS$(X,\mathcal{C}_2;Z,\mathcal{C}_1^\perp)$ code, where $\mathcal{C}_2 = \langle\bm{1} \rangle=$ RM(0,4) $\subset \mathcal{C}_1 =$ RM(1,4) and $\mathcal{C}_1^\perp =$ RM(2,4) $\subset \mathcal{C}_2^\perp =$ RM(3,4) (see \cite{macwilliams1977theory} for more details of classical Reed-Muller codes). The signs of all stabilizers are positive. 
		We know from \cite[Theorem 19]{Opt} that the code space is fixed by transversal $\sqrt{T}$ ($\frac{\pi}{2^4}$ $Z$-rotation), and direct calculation shows that the corresponding logical operator is CCC$Z$ up to some local Pauli corrections. 
		We first verify invariance under transversal $T$ by checking the sufficient conditions given in Remark \ref{rem:arrive Transversal T thm}.
		
		The $[[16, 4, 2]]$ code has a single non-zero $X$-stabilizer $\bm{a}$ = $\bm{1}$, with even weight, and a single subcode $\mathcal{B}(\bm{a}) = \mathcal{C}_1^\perp =$ RM(2,4). This subcode contains a self-dual code, denoted RM($1.5,4$), which is generated by $\bm{1}$, all the degree one monomials, and half of the degree two monomials, i.e., $x_1x_2,x_1x_3,x_1x_4$. 
		Since the weights in RM($1.5,4$) are 0, 4, 8, 12, and 16, we have $ \imath^{w_H(\bm{v})} = 1$ for all $\bm{v} \in$ RM($1.5,4$). This matches the signs specified in the definition of the code above. Hence, the $ [[16, 4, 2]]$ code satisfies the sufficient conditions for invariance under transversal $T$. We note that the logical operator induced by transversal $T$ is the identity (obtained by applying CCC$Z$ twice).
		
		Finally, we verify invariance under transversal $\sqrt{T}$ by checking the first of the trigonometric conditions given in Theorem \ref{thm:transversal_Z_rot}. The weight distribution of $ \mathrm{RM}(2,4)$ is given by
		\begin{equation} \label{weight_en_RM_2_4}
		P(x) = 1+ 140x^4 + 448x^6+870x^8+448x^{10}+140x^{12}+x^{16}.
		\end{equation}
		Let $\alpha_4 = \tan\frac{2\pi}{2^4} =\tan\frac{\pi}{8} $. Since $ (\sec\theta)^2 = 1+ \left(\tan\theta\right)^2$ and $\epsilon_{\bm{v}}=1$, for all $\bm{v}\in \mathcal{B}(\bm{a})$, we have
		\begin{align}
		&\sum_{\bm{v}\in \text{RM}(2,4)} \epsilon_{\bm{v}}\left(\imath \alpha_4\right)^{w_H\left(\bm{v}\right)} - \left(1 + \alpha_4^2 \right)^8  \nonumber\\
		&=
		\left( \imath \alpha_4\right)^0 + 140 \left( \imath \alpha_4 \right)^4 + 448 \left( \imath \alpha_4\right)^6 + 870 \left( \imath \alpha_4 \right)^8 \nonumber\\
		&~~~+ 448 \left( \imath \alpha_4\right)^{10} + 140 \left( \imath \alpha_4\right)^{12} + \left( \imath \alpha_4\right)^{16} - \left(1+\alpha_4^2 \right)^{8} \nonumber\\
		&
		= -8\alpha_4^2(1-\alpha_4)^2(1+\alpha_4)^2(\alpha_4^2+2\alpha_4-1)^2(\alpha_4^2-2\alpha_4-1)^2.
		\end{align}
		The first trigonometric condition is satisfied since $\alpha_4 = \sqrt{2}-1$ is a root of $x^2+2x-1 = 0$. We verified the second condition directly using MATLAB for each nonzero coset representative in $\F_2^{16} / \mathcal{B}(\bm{a})$ and it 
		is also implicit in \cite[Theorem 19]{Opt}.
		
	\end{example}
    Remark \ref{rem:arrive Transversal T thm} motivates the following extension to Lemma \ref{all_one_iff}.
	
	\begin{corollary}\label{extend_iff_div}
		Let $\mathcal{C}$ be a binary linear code with block length $m$ where all codewords have even weight. Suppose that 
		\begin{equation}\label{extend_iff_div_assump}
		\sum_{\bm{v}\in \mathcal{C}} \epsilon_{\bm{v}}\left(\imath\tan\frac{2\pi}{2^l}\right)^{w_H(\bm{v})} = \left(\sec\frac{2\pi}{2^l}\right)^{m},
		\end{equation}
		where $\epsilon: \mathcal{C} \to \{\pm 1\}$ is a character of the additive group $\mathcal{C}$.
		\begin{enumerate}
			\item If $\epsilon_{\bm{v}}=1$ for all $\bm{v}\in \mathcal{C}$, then $2^l$ divides $(m-2w_H(\bm{w}))$ for all $\bm{w}\in \mathcal{C}^\perp$.
			\item If $\epsilon_{\bm{v}}\neq1$ for all $\bm{v}\in \mathcal{C}$, and if $\mathcal{B} = \{\bm{v}\in \mathcal{C}: \epsilon_{\bm{v}} = 1\}$, then $2^l$ divides $(m-2w_H(\bm{w}))$ for all $\bm{w}\in \mathcal{B}^\perp \setminus \mathcal{C}^\perp$.
		\end{enumerate}
	\end{corollary}
	\begin{IEEEproof}
		Part (1) follows from Lemma \ref{all_one_iff}.
		
		To prove part (2), rewrite (\ref{extend_iff_div_assump}) as 
		\begin{align}
		P[\mathcal{B}]-P[\mathcal{C}\setminus \mathcal{B}] 
		=& \sum_{\bm{v}\in \mathcal{B}}\left(\cos\frac{2\pi}{2^l}\right)^{m-w_H(\bm{v})} \left(\imath\sin{\frac{2\pi}{2^l}}\right)^{w_H(\bm{v})}\nonumber \\
		-& \sum_{\bm{v}\in \mathcal{C}\setminus \mathcal{B}} \left(\cos\frac{2\pi}{2^l}\right)^{m-w_H(\bm{v})}\left(\imath\sin{\frac{2\pi}{2^l}}\right)^{w_H(\bm{v})}\nonumber \\
		=& 1
		\end{align}
		Recall the notations we used in the proof of Lemma \ref{all_one_iff} that $t_{+} = \cos\frac{2\pi}{2^l}+\imath\sin{\frac{2\pi}{2^l}}$ and $t_{-}= \cos\frac{2\pi}{2^l}-\imath\sin{\frac{2\pi}{2^l}}$. Since $\bm{1} \in \mathcal{C}^\perp \subset \mathcal{B}^\perp$, we may apply the MacWilliams Identities to obtain
		\begin{align}
		P[\mathcal{B}] + P[\mathcal{C}\setminus \mathcal{B}]
		&= \sum_{\bm{v}\in \mathcal{C}} \left(\cos\frac{2\pi}{2^l}\right)^{m-w_H(\bm{v})}\left(\imath\sin{\frac{2\pi}{2^l}}\right)^{w_H(\bm{v})}  \\
		& =  \frac{1}{|\mathcal{C}^\perp|}P_{\mathcal{C}^\perp}\left(t_{+},t_{-}\right)\\
		&= \frac{1}{|\mathcal{C}^\perp|}\sum_{\bm{w}\in \mathcal{C}^\perp}\cos\left(\frac{2\left(m-2w_H\left(\bm{w}\right)\right)\pi}{2^l}\right).  \label{s_B+s_R}
		\end{align} 
		Note that $\mathcal{B}\subset \mathcal{C}$ is a subspace of index $2$. Since $|\mathcal{B}^\perp| = 2|\mathcal{C}^\perp|$, we may apply the MacWilliams Identities to $P_{\mathcal{B}}\left(\cos\frac{2\pi}{2^l},i\sin\frac{2\pi}{2^l}\right) $ and obtain
		\begin{align}\label{s_R}
		P[\mathcal{B}]
		&= \frac{1}{|\mathcal{B}^\perp|}P_{\mathcal{B}^\perp}\left(t_{+},t_{-}\right) \nonumber\\
		&=\frac{1}{2|\mathcal{C}^\perp|}\sum_{\bm{w}\in \mathcal{B}^\perp}\cos\left(\frac{2\left(m-2w_H\left(\bm{w}\right)\right)\pi}{2^l}\right). 
		\end{align}
		Combining equations (\ref{s_B+s_R}) and (\ref{s_R}) gives
		\begin{align}
		1
		&=P[\mathcal{B}] - P[\mathcal{C} \setminus \mathcal{B}]
		= 2P[\mathcal{B}] - (P[\mathcal{B}]+ P[\mathcal{C}\setminus \mathcal{B}] ) \nonumber \\
		&= \frac{1}{|\mathcal{C}^\perp|}\sum_{\bm{w}\in \mathcal{B}^\perp \setminus \mathcal{C}^\perp}\cos\left(\frac{2\left(m-2w_H\left(\bm{w}\right)\right)\pi}{2^l}\right) .  \label{half_one_iff_last}
		\end{align}
		We complete the proof by observing that each term in \eqref{half_one_iff_last} must contribute 1 to the sum.
	\end{IEEEproof}
	\begin{remark} \normalfont \label{rem:Krawtchouk}
		If $m \neq 0 \pmod{2^{l}}$, then since $\bm{0} \in \mathcal{C}^\perp$, it must be case 2 of Corollary \ref{extend_iff_div} that applies. This is always the case when $2^l>m$. We must have $w_H(\bm{v}) = {m}/{2}$ for all $\bm{v} \in \mathcal{B}^\perp \setminus \mathcal{C}^\perp$, and we remark that if we expand the MacWilliams Identities using Krawtchouk polynomials~\cite{macwilliams1977theory}, then we can can show that there exist at least $m/2$ codewords in $\mathcal{C}$ with Hamming weight $2$.
	\end{remark}

	By setting $\mathcal{C} = \mathcal{B}(\bm{a})$ in Theorem~\ref{thm:transversal_Z_rot}, we see that the scenario $2^l > w_H(\bm{a})$ applies whenever we require that Theorem~\ref{thm:transversal_Z_rot} holds for all $l \geq 2$. Thus, the observation using Krawtchouk polynomials implies the existence of a large set of weight $2$ $Z$-stabilizers in the code. This motivates the study of stabilizers groups with such structure, which we embark upon next, noting that existence is proved in Theorem \ref{thm:trans_Z_all_l}.

	\section{Weight Two $Z$-Stabilizers}
	\label{sec:weight_2_Zs}

	We begin this section by examining the structure of a stabilizer group $\mathcal{S}$ that contains weight $2$ $Z$-stabilizers. Later in this section we show (in the proof of necessity in Theorem~\ref{thm:trans_Z_all_l}) that if a stabilizer code $\mathcal{V}(\mathcal{S})$ is preserved by the transversal $\pi/2^l$ $Z$-rotation for all $l\ge 2$, then $\mathcal{S}$ contains a large number of weight $2$ $Z$-stabilizers. 
	
	Let $\bm{e_i},~ i=1,2,\dots,n$ be the standard basis of $\F_2^n$. Recall the graph with vertex set 
	\begin{equation}
	    \Gamma = \bigcup_{\epsilon E(\bm{a},\bm{b}) \in \mathcal{S}} \mathrm{supp}(\bm{a}),
	\end{equation}
	where vertices $i$ and $j$ are joined if $\epsilon E(\bm{0},\bm{e_i} \oplus \bm{e_j})\in S$ for some $\epsilon\in\{\pm 1\}$. Recall that we denote the connected components of the graph by $\Gamma_1,\cdots,\Gamma_{t}$, and set $N_k = |\Gamma_k|$ for $k=1,2,\cdots,t$.

	\begin{lemma}\label{completeness}
		Each component $\Gamma_{k}$, $k=1,2,\cdots,t$ is a complete graph.
	\end{lemma}
	\begin{IEEEproof}
		If a path $r_0,r_1,\cdots,r_j$ connects vertices $r_0$ and $r_j$, then $r_0$ is joined to $r_j$ since
        \begin{IEEEeqnarray*}{rCl+x*}
        \pm E\left(\bm{0},\bm{e_{r_0}}\oplus \bm{e_{r_j}}\right) & = & \prod_{i=0}^{j-1}\left[\pm E\left(\bm{0},\bm{e_{r_i}}\oplus \bm{e_{r_{i+1}}}\right)\right].
		&
		\IEEEQEDhere
		\end{IEEEeqnarray*}
	\end{IEEEproof}

	This implies that the $Z$-stabilizers corresponding to $\Gamma_k$ are given by all length $N_k$ vectors of even weight, i.e., the $[N_k, N_k-1, 2]$ single parity check code. 
	Henceforth, we denote the $[m, m-1, 2]$ single parity check code of any length $m$ by $\mathcal{W}$.
	Theorem~\ref{thm:transversal_Z_rot} forces us to consider all $Z$-stabilizers $\mathcal{B}(\bm{a})$ supported on the $X$-component $\bm{a}$ of some stabilizer $\epsilon E(\bm{a},\bm{b})$.
	The next observation shows that $\bm{a}$ either has full support or no support on a given $\Gamma_k$.
	Together with the above result, this means that each $\Gamma_k$ either contributes $(N_k-1)$ dimensions worth of $Z$-stabilizers or nothing at all to $\mathcal{B}(\bm{a})$.
	This suggests that we split the sum that appears in Theorem~\ref{thm:transversal_Z_rot} in terms of smaller sums over the $\Gamma_k$'s lying within the support of $\bm{a}$.
	Indeed, we are building up towards such an argument in Theorem~\ref{thm:trans_Z_all_l}.

	Given $\bm{v}\in \F^n_2$, let $\bm{v_k} = \bm{v} \big|_{\Gamma_{k}} \in \mathbb{F}_2^{N_k}$ be the restriction of $\bm{v}$ to $\Gamma_{k}$ for $k=1,\dots,t$.
	
	\begin{lemma}\label{xstabilizerconstant}
		If $\pm E(\bm{a},\bm{b})$ is a stabilizer in $\mathcal{S}$, then $\bm{a_k} = \bm{0}$ or $\bm{1}$. 
	\end{lemma}
	\begin{IEEEproof}
	$\pm E(\bm{a},\bm{b})$ commutes with $	\pm E\left(\bm{0},\bm{e_{r_i}}\oplus \bm{e_{r_j}}\right) $ for all $i,j \in \Gamma_k$.
	\end{IEEEproof}
    The $Z$-stabilizers supported on $\Gamma_k$ take the form $(-1)^{\bm{y_k} \bm{v}^T} E(\bm{0},\bm{v})$, where $\bm{v}$ is a vector of even weight supported on $\Gamma_k$. Here $\bm{y}_k$ is a fixed binary vector supported on $\Gamma_k$. We now investigate trigonometric identities satisfied by the weights in these component codes $\mathcal{W}$ representing $Z$-stabilizers from $\Gamma_k$. 
	
	\begin{lemma}
    Let $\mathcal{W}$ be the $[m,m-1]$ code consisting of all vectors with even weight, and let $\epsilon_{\bm{v}} = (-1)^{\bm{v}\bm{y}^T}$ be a character on $\mathcal{W}$. Then 
\begin{equation}\label{half_one_P_W}
\sum_{\bm{v}\in \mathcal{W}} \epsilon_{\bm{v}}\left(\imath\tan\frac{2\pi}{2^l}\right)^{w_H(\bm{v})} = \cos \gamma \cdot\left(\sec\frac{2\pi}{2^l}\right)^m,
\end{equation}
where $\gamma = \frac{2\pi\left(M-2w_H\left(\bm{y}\right)\right)}{2^l}$.
\end{lemma}
\begin{IEEEproof}
		If $\epsilon$ is the trivial character, then $\bm{y}={\bm{0}}$, and we have
		\begin{equation}
		\frac{\sum_{\bm{v}\in \mathcal{W}}\left(\imath \tan\frac{2\pi}{2^l}\right)^{w_H(\bm{v})} }{\left(\sec\frac{2\pi}{2^l}\right)^{m}}=P\left[\mathcal{W}\right].
		\end{equation}
		We apply the MacWilliams Identities to obtain
		\begin{align}
		P\left[\mathcal{W}\right]
		& = \frac{1}{|\mathcal{W}^\perp|}P_{\mathcal{W}^\perp}\left(\cos\frac{2\pi}{2^l}+\imath\sin\frac{2\pi}{2^l},\cos\frac{2\pi}{2^l}-\imath\sin\frac{2\pi}{2^l}\right) \nonumber \\
		& = \frac{1}{|\mathcal{W}^\perp|}P_{\mathcal{W}^\perp}\left(e^{\imath\frac{2\pi}{2^l}},e^{-\imath\frac{2\pi}{2^l}}\right) \nonumber\\
		& = \cos\frac{2\pi m}{2^l}, 
		\end{align}
		which means
				\begin{equation}\label{P_W}
		\sum_{\bm{v}\in \mathcal{W}}\left(\imath \tan\frac{2\pi}{2^l}\right)^{w_H(\bm{v})} =
		\cos\frac{2\pi M}{2^l}\left(\sec\frac{2\pi}{2^l}\right)^{m}.
		\end{equation} 
	
	If $\epsilon$ is a non-trivial character, then there exists $\bm{y}\in\F^m_2$ with $\bm{y} \neq \bm{0}$ or $ \bm{1}$ such that 
	\begin{equation}
	\mathcal{B}= \{\bm{v} \in W : \epsilon_{\bm{v}}=1\} = \langle \bm{1},\bm{y} \rangle^\perp,
	\end{equation}
	and
	\begin{equation}
	\mathcal{B}^\perp = \langle \bm{1},\bm{y} \rangle = \{\bm{0}, \bm{1},\bm{y},\bm{1} \oplus \bm{y} \}.
	\end{equation}
	Note that $|\mathcal{B}| =\frac{|\mathcal{W}|}{2}$ and $|\mathcal{B}^\perp| = 2|\mathcal{W}^\perp|$. We rewrite 
	\begin{align}
	\sum_{\bm{v}\in \mathcal{W}} \epsilon_{\bm{v}} &\left(\imath\tan\frac{2\pi}{2^l}\right)^{w_H(\bm{v})} \nonumber\\
	& = \sum_{\bm{v}\in \mathcal{B}} \left(\imath\tan\frac{2\pi}{2^l}\right)^{w_H(\bm{v})} -  \sum_{\bm{v}\in \mathcal{W} \setminus \mathcal{B}} \left(\imath\tan\frac{2\pi}{2^l}\right)^{w_H(\bm{v})}\\
	& =  2\sum_{\bm{v}\in \mathcal{B}} \left(\imath\tan\frac{2\pi}{2^l}\right)^{w_H(\bm{v})} -  \sum_{\bm{v}\in \mathcal{W} } \left(\imath\tan\frac{2\pi}{2^l}\right)^{w_H(\bm{v})},
	\end{align}
	so that 
	\begin{equation}
	\frac{\sum_{\bm{v}\in \mathcal{W}}\epsilon_{\bm{v}}\left(\imath \tan\frac{2\pi}{2^l}\right)^{w_H(\bm{v})} }{\left(\sec\frac{2\pi}{2^l}\right)^{m}} =2P\left[\mathcal{B}\right]-P\left[\mathcal{\mathcal{W}}\right].
	\end{equation}
	We apply the MacWilliams Identities to obtain
	\begin{align}
	P\left[\mathcal{B}\right]
	&=\frac{1}{|\mathcal{B}^\perp|}P_{\mathcal{B}^\perp}\left(e^{\imath\frac{2\pi}{2^l}},e^{-\imath\frac{2\pi}{2^l}}\right) \nonumber\\
	&=\frac{1}{2}\left[\cos\frac{2\pi m}{2^l}+\cos\frac{2\pi (m-2w_H(\bm{y}))}{2^l}\right].
	\end{align}
	We combine with \eqref{P_W} to obtain
	\begin{equation}
	2P\left[\mathcal{B}\right]-P\left[\mathcal{W}\right] = \cos\frac{2\pi \left(m-2w_H\left(\bm{y}\right)\right)}{2^l}
	\end{equation}
	as required.
\end{IEEEproof}

	When $\mathcal{B}(\bm{a}) = \mathcal{W}$, the second trigonometric identity in Theorem \ref{thm:transversal_Z_rot} becomes a sum over all odd weight vectors $\left(\mathbb{F}_2^m \setminus \mathcal{W}\right)$. The character $\epsilon$ is given by $\epsilon_{\bm{v}} = (-1)^{\bm{v} \bm{y}^T}$ for some $\bm{y} \in \F_2^m$ and we extend the domain of $\epsilon$ from $\mathcal{W}$ to $\F_2^m$. If $\epsilon$ is trivial, then 
	\begin{align}
	\frac{\sum_{\bm{v}\in \F^m_2 \setminus \mathcal{W}} \epsilon_{\bm{v}}\left(\imath\tan\frac{2\pi}{2^l}\right)^{w_H(\bm{v})}}{\left(\sec\frac{2\pi}{2^l}\right)^m} 
	&= P\left[\F^m_2\setminus \mathcal{W} \right] \nonumber\\
	&= P\left[\F^m_2\right] - P\left[\mathcal{W}\right].
	\end{align}
	We apply the MacWilliams Identities to obtain 
	\begin{align}
	P\left[\F^m_2 \right]
	& = P_{\langle \bm{0} \rangle} \left(e^{\imath\frac{2\pi}{2^l}}, e^{-\imath\frac{2\pi}{2^l}}\right)\\
	& = \left(e^{\imath\frac{2\pi}{2^l}}\right)^{m-0}\left(e^{\imath\frac{2\pi}{2^l}}\right)^0\\
	& = \cos\frac{2\pi m}{2^l} + \imath\sin\frac{2\pi m}{2^l} \label{P_F}.
	\end{align}
	It now follows from equation (\ref{P_W}) that 
	\begin{equation}  \label{P_F_minus_W}
	P\left[\F^m_2\right] - P\left[\mathcal{W}\right]
	= \imath\sin\frac{2\pi m}{2^l} 
	= \imath\sin\frac{2\pi \left(m-2w_H\left(\bm{0}\right)\right)}{2^l}. 
	\end{equation}
	If $\epsilon$ is non-trivial, let $\mathcal{B}'=\{x\in\F^m_2| \epsilon_{x}=1\}$. If $\mathcal{B}'=\mathcal{W}$, then
	\begin{align} \label{P_F_minus_W_B=1}
	\frac{\sum_{\bm{v}\in\F^m_2 \setminus \mathcal{W}} \epsilon_{v} \left(\imath\tan \frac{2\pi}{2^l}\right)^{w_H(\bm{v})}}{\left(\sec\frac{2\pi}{2^l}\right)^m} 
	&= -\imath\sin\frac{2\pi m}{2^l} \nonumber\\
	&= \imath\sin\frac{2\pi (m-2w_H(\bm{1}))}{2^l}.
	\end{align}
	Note that since $ \langle \bm{y}\rangle  \subseteq \langle \bm{1}, \bm{y}\rangle = \mathcal{B}^\perp $, we have $B \subseteq \bm{y}^\perp $. It remains to consider the case where $\epsilon$ is non-trivial and $\mathcal{B}'\neq \mathcal{W}$. Here $\mathcal{B}'=\bm{y}^\perp$ where $\bm{y}\neq \bm{1}$. 

	\begin{lemma}
		\label{halfonesomeodd} 
	 Let $\mathcal{W}$ be the $[m,m-1]$ code consisting of all vectors with even weight. Let $ \epsilon_{\bm{v}} = (-1)^{\bm{v} \bm{y}^T}$, let $\mathcal{B}= \{\bm{v} \in \mathcal{W}|\epsilon_{\bm{v}}=1\} = \langle \bm{1},\bm{y} \rangle^\perp$, and let $\mathcal{B}'=\{\bm{x}\in\F^m_2| \epsilon_{\bm{x}}=1\}$.Then
	\begin{equation}\label{halfonesomeodd_conclusion}
	\sum_{\bm{v}\in \F^{m}_2 \setminus {\mathcal{W}}}\epsilon_{\bm{v}} \left(\imath \tan\frac{2\pi}{2^l}\right)^{w_H(\bm{v})} =\imath\sin\gamma \cdot \left(\sec\frac{2\pi}{2^l}\right)^{m},
	\end{equation}
	where $\gamma = \frac{2\pi\left(m-2w_H\left(\bm{y}\right)\right)}{2^l} $.
\end{lemma}
\begin{IEEEproof}
	See Appendix \ref{sec:proof_of_odd_wt}.
\end{IEEEproof}

We now consider a stabilizer code $\mathcal{V}(\mathcal{S})$ that is preserved by $\pi/2^l$ $Z$-rotation for all $l\ge 2$. The sign $\epsilon_{\bm{v}}$ of the $Z$-stabilizer $\epsilon_{\bm{v}}E(\bm{0},\bm{v})$ is given by $\epsilon_{\bm{v}}=(-1)^{\bm{y}\bm{v}^T}$, and we let $\bm{y_k}=\bm{y}\big|_{\Gamma_k}$ be the restriction of the binary vector $\bm{y}$ to $\Gamma_k$. Given $\epsilon E(\bm{a},\bm{b}) \in \mathcal{S}$ with $\bm{a}\neq \bm{0}$, we now investigate the trigonometric conditions satisfied by $Z$-stabilizers supported on $\mathrm{supp}(\bm{a})$. We first show that $\mathrm{supp}(\bm{a})$ is the disjoint union of components $\Gamma_k \subseteq \mathrm{supp}(\bm{a})$. We then glue together the trigonometric conditions satisfied by the $Z$-stabilizers supported on these components $\Gamma_k$.


\addtocounter{theorem}{-9}
	\begin{theorem}\label{thm:trans_Z_all_l}
	A transversal $\pi/2^l$ $Z$-rotation preserves the stabilizer code for all $l\ge 2$ if and only if for every $\epsilon E(\bm{a},\bm{b}) \in \mathcal{S}$ with $\bm{a}\neq \bm{0}$, 
	\begin{enumerate}
	    \item $\mathrm{supp}(\bm{a})$ is the disjoint union of components $\Gamma_k \subseteq \mathrm{supp}(\bm{a})$,
	    \item $N_k$ is even and $w_H(\bm{y_k})= {N_k}/{2}$ for all $k$ such that $\Gamma_k \subseteq \mathrm{supp}(\bm{a})$.
	\end{enumerate}
	\end{theorem}
	\addtocounter{theorem}{+8}
\begin{IEEEproof}[Proof of Necessity]
    First, we need to show that the hypothesis implies the presence of many weight $2$ $Z$-stabilizers, and hence that the discussion of $\Gamma_k$ is material.
    Though we remarked on their presence in Remark~\ref{rem:Krawtchouk}, we will see in this proof that such a structure is revealed by the trigonometric conditions in Theorem~\ref{thm:transversal_Z_rot} itself.
    For now, we begin by assuming their presence and introducing related quantities.
    
	We divide the  weight $2$ $Z$-stabilizers in $\Gamma_k$ into two classes of sizes ${P}_k$ and ${Q}_k$ where ${P}_k=|\{\bm{v}\in \F_2^{|\Gamma_k|}: w_H(\bm{v})=2 \text{ and } \epsilon_{\bm{v}}=1 \}|$ and  ${Q}_k=|\{\bm{v}\in \F_2^{|\Gamma_k|} : w_H(\bm{v})=2 \text{ and } \epsilon_{\bm{v}}=-1 \}|$. Setting $w_H(\bm{y_k})=s$, we have
			\begin{align}
			{Q}_k-{P}_k&=\binom{s}{1}\binom{N_k-s}{1}-\left(\binom{s}{2}+\binom{N_k-s}{2} \right)\\
			&=-2\left(s-\frac{N_k}{2} \right)^2+\frac{N_k}{2}. \label{eqn:claim}
			\end{align}
			Thus, ${Q}_k-{P}_k \le \frac{N_k}{2}$, and equality holds if and only if  $w_H(\bm{y_k})=\frac{N_k}{2}$.
			Theorem~\ref{thm:transversal_Z_rot} implies all $w_H(\bm{a})$ are even and
				\begin{equation}\label{eqn:sec}
				\sum_{\bm{v}\in \mathcal{B}(\bm{a})} \epsilon_{\bm{v}}\left(\imath \tan\theta \right)^{w_H(\bm{v})} = \left(\sec\theta\right)^{w_H(\bm{a})}=(1+(\tan\theta)^2)^{\frac{w_H(\bm{a})}{2}} 
				\end{equation}
		for all $\theta=\frac{\pi}{2^l} \text{ with } l\ge 2$. Let $\mathcal{B}_{2j}(\bm{a})=\{\bm{z}\in \mathcal{B}(\bm{a})| w_H(\bm{z})=2j \}$. We have 
		\begin{equation}\label{eqn:zero_poly}
		\sum_{j=0}^{\frac{w_H(\bm{a})}{2}}\sum_{\bm{v}\in \mathcal{B}_{2j}(\bm{a})} \epsilon_{\bm{v}}(-1)^j\left( \tan\theta \right)^{2j} =\left(1+(\tan\theta)^2\right)^{\frac{w_H(\bm{a})}{2}}.
		\end{equation}
		for all $\theta=\frac{\pi}{2^l} \text{ with } l\ge 2$. 
		Since a finite degree polynomial (in $\left(\tan\theta\right)^2$) cannot have infinitely many roots $\left(\tan\frac{\pi}{2^l}\right)^2$, it must be identically zero and we may equate the coefficients of $\left(\tan\theta\right)^2$ to obtain
		\begin{equation}
		\frac{w_H(\bm{a})}{2}=\sum_{\bm{v}\in \mathcal{B}_{2}(\bm{a})} \epsilon_{\bm{v}}\cdot (-1)= \sum_{k: \Gamma_k\subseteq \mathrm{supp}(\bm{a})} (Q_k-P_k ).
		\end{equation}
		Note that this observation has established the presence of weight $2$ vectors in $\mathcal{B}(\bm{a})$, as we intended.
		It follows from \eqref{eqn:claim} that 
		\begin{equation}\label{eqn:ineq_eq}
		\frac{w_H(\bm{a})}{2}\leq \sum_{k: \Gamma_k\subseteq \mathrm{supp}(\bm{a})} \frac{N_k}{2}\leq \frac{w_H(\bm{a})}{2}.
		\end{equation}
	Therefore equality holds in \eqref{eqn:ineq_eq} and ${Q}_k-{P}_k=\frac{N_k}{2}$ for all $k$ such that $\Gamma_k\subseteq \mathrm{supp}(\bm{a})$, which completes the proof. 
	\linebreak
	
	\textit{Proof of Sufficiency.}
		 Let $\mathcal{W}_k^0$ be the $[N_k,N_k-1]$ single-parity-check code and let $\mathcal{W}_k^1=\F_2^{N_k}\setminus \mathcal{W}_k^0$. Let $\mathcal{W}(\bm{r}) = \bigoplus_{k: \Gamma_k\subseteq \mathrm{supp}(\bm{a})}\mathcal{W}_k^{{r}_k}$, where $\bm{r}\in \F_2^{|\{k:\Gamma_k\subseteq \mathrm{supp}(\bm{a})\}|}$ and $r_{k}$ is the entry of $\bm{r}$ corresponding to $\Gamma_k$. Then, for all $\bm{r}$,
		\begin{align}\label{main_thm_partition}
		&\sum_{\bm{v}\in \mathcal{W}(\bm{r})} \epsilon_{\bm{v}}\left(\imath \tan \frac{2\pi}{2^l}\right)^{w_H(\bm{v})} =  \prod_{\substack{ k \\ \Gamma_k\subseteq \mathrm{supp}(\bm{a}) }}  f_{k}(r_k),
		\end{align}
		where 
		\begin{equation}\label{eqn:inner_sum}
		 f_{k}(\delta) = \sum_{\bm{\eta} \in \mathcal{W}_k^{\delta}} (-1)^{\bm{y_k}\bm{\eta}^T} \left(\imath \tan\frac{2\pi}{2^l}\right)^{w_H(\bm{\eta})}, \text{ for } \delta\in\{0,1\}.
		\end{equation}
		
		Here, $\bm{y_k}=\bm{y}\big|_{\Gamma_k}$ be the restriction of the character vector $\bm{y}$ to $\Gamma_k$. Let $\gamma=\frac{2\pi\left(N_k-2w_H\left(\bm{y_k}\right)\right)}{2^l}$. We apply \eqref{P_W} and \eqref{halfonesomeodd_conclusion} to simplify \eqref{eqn:inner_sum} as
		\begin{align}
		 f_{k}(\delta)
		& =\left\{ \begin{array}{lc}
		\cos\gamma \cdot \left(\sec\frac{2\pi}{2^l}\right)^{N_k} ~~ \text{ if } \delta=0, \\
		\imath\sin\gamma \cdot \left(\sec\frac{2\pi}{2^l}\right)^{N_k} ~ \text{ if } \delta=1,
		\end{array}\right. \nonumber \\
		& = \left\{ \begin{array}{lc}
		\left(\sec\frac{2\pi}{2^l}\right)^{N_k} & \text{ if } \delta=0, \\
		0 & \text{ if } \delta=1.
		\end{array}\right. \label{simplified_general_form}
		\end{align}
		Therefore, the summation \eqref{main_thm_partition} is nonzero if only if $\bm{r}=\bm{0}$ (i.e. summing over $\mathcal{W}(\bm{0})$).

		To show the first trigonometric identity in Theorem \ref{thm:transversal_Z_rot}, we note that $\mathcal{B}(\bm{a})\supset \mathcal{W}(\bm{0})$. Then, for all $l\ge 3$
		\begin{align} \label{main_theorem_first_condition}
		\sum_{\bm{v}\in \mathcal{B}(\bm{a})} \epsilon_{v}\left(\imath \tan\frac{2\pi}{2^l}\right)^{w_H(\bm{v})} 
		&=\sum_{\bm{v}\in \mathcal{W}} \epsilon_{\bm{v}}\left(\imath \tan \frac{2\pi}{2^l}\right)^{w_H(\bm{v})} \nonumber\\
		& = \prod_{\substack{k \\ \Gamma_k\subseteq \mathrm{supp}(\bm{a}) }} \left(\sec\frac{2\pi}{2^l}\right)^{N_k} \nonumber\\
		& = \left(\sec\frac{2\pi}{2^l}\right)^{w_H(\bm{a})}.
		\end{align}
		To verify the second condition, let $\bm{\omega}\in \mathcal{O}(\bm{a}) = \F_2^{w_H(\bm{a})}\setminus \mathcal{B}(\bm{a})$ and we change variables to $\bm{\beta} = \bm{v}\oplus \bm{\omega}$ and 
		$\bm{\omega}$ on the right hand side 
		 (note that we have extended the $\epsilon_{\bm{v}}$ to all binary vectors). Since $\mathcal{W}(\bm{0})$ is not contained in any nontrival coset of $\mathcal{B}(\bm{a})$, we have
		 \begin{align}\label{eqn:change_variable}
		 &\sum_{\bm{v}\in \mathcal{B}(\bm{a})} \epsilon_{\bm{v}}\left(\imath \tan \frac{2\pi}{2^l}\right)^{w_H(\bm{v}\oplus \bm{\omega})} \nonumber \\
		 &= \epsilon_{\bm{\omega}}\sum_{\bm{\beta} \in \bm{\omega} \oplus \mathcal{B}(\bm{a})} \epsilon_{\bm{\beta}}\left(\imath \tan \frac{2\pi}{2^l}\right)^{w_H(\bm{\beta})} = 0,
		 \end{align}
		 for all $l\ge 3$ and $\bm{\omega}\neq \bm{0}$. 
		 \end{IEEEproof}

We now use the two conditions in Theorem \ref{thm:trans_Z_all_l} to show that if a CSS code is oblivious to coherent noise, then it is a constant excitation code. 
\begin{corollary}\label{coro:ce_is_nece}
     A CSS code is oblivious to coherent noise if and only if it is a constant excitation code. 
     
     If the CSS code is error-detecting ($d>1$) then the weights in different cosets of the $X$-stabilizers are identical. 
\end{corollary}
\begin{IEEEproof}
    Consider an $[[n,k,d]]$ CSS($X,\mathcal{C}_2;Z,\mathcal{C}_1^\perp$) code with a fixed character vector $\bm{y}$ for $Z$-stabilizers. If $\bm{w}$ is a coset representative for $\mathcal{C}_2$ in $\mathcal{C}_1$, then $\bm{w}\perp \mathcal{C}_1^\perp$ so $\bm{w}\big|_{\Gamma_k} = \bm{0} \text{ or } \bm{1}.$ If $\bm{x}\in\mathcal{C}_2$, then by Lemma \ref{xstabilizerconstant}, we have $\bm{x}\big|_{\Gamma_k} = \bm{0} \text{ or } \bm{1}$ for all $k$. Theorem \ref{thm:trans_Z_all_l} implies $w_H(\bm{y_k}) = \frac{|\Gamma_k|}{2}$ for all $k$, where $\bm{y_k} = \bm{y}\big|_{\Gamma_k}$. 
    Since $(\bm{w} \oplus \bm{x}) = \bm{0} \text{ or } \bm{1}$ on any $\Gamma_k$, adding $\bm{y_k}$ to the sum either leaves $\bm{y_k}$ unchanged or just flips all entries of $\bm{y_k}$.
    In both cases, the Hamming weight of the sum $(\bm{w}\oplus\bm{x}\oplus\bm{y})$ is exactly $\frac{|\Gamma_k|}{2}$ on any $\Gamma_k$.
    If $\Gamma=\bigcup_{k=1}^{t}\Gamma_k$, then 
    \begin{equation}
        w_H(\bm{w}\oplus\bm{x}\oplus\bm{y}\big|_{\Gamma}) = \frac{\sum_{k=1}^{t}|\Gamma_k|}{2}.
    \end{equation}
    If $V= \{1,2,\dots,n\} \setminus \Gamma$, then the first condition in Theorem \ref{thm:trans_Z_all_l} implies that $ w_H(\bm{x}\big|_{V}) = \bm{0}$, so that for fixed $\bm{w}$
        \begin{equation}\label{eqn:ce_split}
             w_H(\bm{w}\oplus\bm{x}\oplus\bm{y}) =  w_H(\bm{w}\oplus\bm{x}\oplus\bm{y}\big|_{\Gamma}) +  w_H(\bm{w}\oplus\bm{x}\oplus\bm{y}\big|_{V}) 
        \end{equation}
        is constant for all $\bm{x} \in \mathcal{C}_2$, and the CSS code is a constant excitation code. The sufficiency follows from the observation that a transversal $\theta$ $Z$-rotation acts as a global phase on a constant excitation code. If the CSS code is error detecting, then for all $i\in V$ there exists $\epsilon_i \in \{\pm 1\}$ such that $\epsilon_i E(\bm{0},\bm{e_i})$ is a $Z$-stabilizer. Hence $\bm{w}\big|_{\bm{v}}=\bm{0}$ for all coset representatives $\bm{w}=\bm{v}G_{\mathcal{C}_1/\mathcal{C}_2}$ of $\mathcal{C}_2$ in $\mathcal{C}_1$. It now follows from \eqref{eqn:ce_split} that $w_H(\bm{w})=\frac{|\Gamma|}{2}+w_H(\bm{y}\big|_{\bm{v}})$ is constant. 
\end{IEEEproof}

	\section{Construction of Quantum Codes oblivious to Coherent Noise }
	\label{sec:construction_method}
	Let $\mathcal{A}_2\subset \mathcal{A}_1$ be two classical codes with length $t$, and let $R_2, R_1$ respectively be the rates of $\mathcal{A}_2, \mathcal{A}_1$. We may construct a $[[t, (R_2-R_1)t, d= \min\{d_{\min}(\mathcal{A}_1),d_{\min}(\mathcal{A}_2^\perp)\}]]$ CSS code by choosing $X$-stabilizers from $\mathcal{A}_2$ and $Z$-stabilizers from $\mathcal{A}_1^\perp$. Let $M\ge 2$ be even, and let $\mathcal{W}$ be the $[M,M-1]$ single parity check code consisting of all vectors with even weight of length $M$. Consider the CSS$(X,\mathcal{C}_2;Z,\mathcal{C}_1^\perp)$ code where
	\begin{equation}
	    \mathcal{C}_2=\mathcal{A}_2\otimes \bm{1}_M,
	\end{equation}
	\begin{equation}
	    \mathcal{C}_1^\perp=\left\{\left(\bm{b}\otimes \bm{e_1}\right) \oplus \bm{w}:\bm{b}\in \mathcal{A}_1^\perp \text{ and } \bm{w}\in \bigoplus_{k=1}^t \mathcal{W} \right\},
	\end{equation}
	and $\bm{1}_M$ is the all-ones vector of length $M$.
	Note that the code $\mathcal{C}_1^\perp$ includes the direct sum of $t$ single-parity-check codes $\mathcal{W}$. We determine signs of elements in $\mathcal{C}_1^\perp$ ($Z$ stabilizers) by choosing a character vector $\bm{y}\in \F_2^{tM}$, and we satisfy condition (2) of Theorem \ref{thm:trans_Z_all_l} by choosing $w_H(\bm{y_k})=M/2$, where $\bm{y_k}=\bm{y}\big|_{\Gamma_k}$. The sign $\epsilon_{\bm{z}}$ of the $Z$-stabilizer $\epsilon_{\bm{z}}E(\bm{0},\bm{z})$ is given by $\epsilon_{\bm{z}}= (-1)^{\bm{y_k}\bm{z}^T}$. The number of logical qubits is 
	\begin{align}
	    &tM-\dim(\mathcal{C}_1^\perp)-\dim(\mathcal{C}_2) \nonumber\\
	    &=tM-t(M-1)-(1-R_1)t-R_2t = (R_2-R_1)t.
	\end{align}
	If $z$ is a vector of minimum weight that is orthogonal to all $X$-stabilizers, then either $\bm{z}$ is a $Z$-stabilizer of $\bm{z}$ is a vector from $\mathcal{A}_2^\perp$ interspersed with zeros. Hence the minimum distance $d$ of the CSS code is at least $\min(d_{\min}(\mathcal{A}_1)M,d_{\min}(\mathcal{A}_2^\perp))$. Thus, we have constructed a CSS code family with parameters  $[[tM,(R_2-R_1)t, \ge \min(d_{\min}(\mathcal{A}_1)M, d_{\min}(\mathcal{A}_2^\perp))]]$, that is oblivious to coherent noise. 
	
	For fixed $M$, if we choose a family CSS codes with finite rate, then the new CSS family also have finite rate but with possible higher distances. If we allow both $M$ and $t$ to grow without bound, then the new CSS family may achieve increased distance but will have vanishing rate.
	\begin{example}
	    We may choose $\mathcal{A}_1 = \F_2^{2L}$, $\mathcal{A}_2$, and $M=2L$ to be the $[2L,2L-1]$ single-parity-check code to obtain the family of $[[4L^2,1,2L]]$ Shor codes. 
	\end{example}
	 The dual-rail inner code \cite{knill2001scheme} is the CSS code determined by the specific stabilizer group $\mathcal{S}=\langle -Z_1Z_2\rangle$. Ouyang \cite{Ouyang-arxiv20b} observed that it was possible to construct a constant excitation code by concatenating an outer stabilizer code with an inner dual-rail code. This is simply because concatenation maps $\ket{0}$ to $\ket{01}$ and $\ket{1}$ to $\ket{10}$. In this case the number of physical qubits doubles. When $M=2$, the construction described above coincides with the dual-rail construction. 
	 However, our approach has shown that \emph{any} CSS code can be made oblivious to coherent noise, without requiring a special stabilizer group as in the original dual-rail construction.
	 In fact, our approach can be extended to any stabilizer code as shown below.
	 
	 	Consider an $[[n,k,d]]$ stabilizer code with generator matrix
	\begin{align}
	G_{\mathcal{S}}=
	\begin{blockarray}{ccc}
	n & n &   \\
	\begin{block}{[c|c]c}
	A  & B & r-l \\
	\cmidrule(lr){1-2}
	  & C & l  \\
	\end{block}
	\end{blockarray}.
	\end{align}
	Here, $r=n-k$, and the matrix $C$ is the generator matrix of the space $\{\bm{z}\in \F_2^n| \epsilon_{\bm{z}}E(0,\bm{z})\in S \}$ (thus the matrix $A$ has full row rank). The stabilizer code derived from our construction has generator matrix
	
	\begin{align}\label{three_part}
	\setlength\aboverulesep{0pt}\setlength\belowrulesep{0pt}
	\setlength\cmidrulewidth{0.5pt}
	G_{\mathcal{S}'}=
	\begin{blockarray}{ccc}
	nM & nM &   \\
	\begin{block}{[c|c]c}
	A\otimes\bm{1}_{M}  & B\otimes \bm{e}_1 & r-l \\
	\cmidrule(lr){1-2}
	  & C\otimes \bm{e}_1 & l  \\
	\cmidrule(lr){1-2}
	 & I_n\otimes W & n(M-1) \\
	\end{block}
	\end{blockarray},
	\end{align}
	where the $(M-1)\times M$ matrix $W$ generates the single-parity-check code. We choose signs of the $n(M-1)$ stabilizers generated by $I_n \otimes W$ so that the new stabilizer code is oblivious to coherent noise.
	
	\begin{theorem}\label{distance}
	    The minimum distance $d'$ of the stabilizer code generated by $G_{\mathcal{S}'}$ satisfies $d\le d'\le Md$. 
	\end{theorem}
	\begin{IEEEproof}
	    Suppose that $(\bm{x},\bm{y})$ is not in the row space of $G_{\mathcal{S}'}$ and $G_{\mathcal{S}'}(\bm{y},\bm{x})^T = 0$. Note that $M\mid w_H(\bm{x})$. We may write 
	    \begin{equation}
	        \bm{x} = \bm{f} \otimes \bm{1}_M \text{ where } \bm{f}\in \F_2^n,
	    \end{equation}
	    and $\bm{y} = (\bm{1}_M \otimes (\bm{w_1},\dots, \bm{w_n}) )\oplus (\bm{g} \otimes \bm{e_1} )\text{ where } \bm{w_i} \in \mathcal{W} \text{ and } \bm{g}\in \F_2^n.$
	    Then 
	    \begin{align}
	    \setlength\aboverulesep{0pt}\setlength\belowrulesep{0pt}
    	\setlength\cmidrulewidth{0.5pt}
    	G_{\mathcal{S}'}(\bm{y},\bm{x})^T=
    	\left[
    	\begin{array}{c|c}
    	A  & B  \\
    	\cmidrule(lr){1-2}
    	  & C  \\
    	\end{array}
    	\right]
    	(\bm{g},\bm{f})^T = 0.
	    \end{align}
	The weight of $(\bm{x},\bm{y})$ is at least the weight of $(\bm{f},\bm{g})$ which is at least $d$, and so $d'\ge d$. Furthermore, there exists a weight $d$ vector $(\bm{u},\bm{v})$ not in the row space of $G_{\mathcal{S}}$ and $G_{\mathcal{S}}(\bm{v},\bm{u})^T = 0$. Then, we have $(\bm{u} \otimes \bm{1}_M,\bm{v} \otimes \bm{e_1})$ is not in the row space of $G_{\mathcal{S'}}$ 
	and $G_{\mathcal{S'}}(\bm{v} \otimes \bm{e_1},\bm{u} \otimes \bm{1}_M)^T = 0$. Hence,
    \begin{IEEEeqnarray*}{rCl+x*}
      d'\le w_H(\bm{u} \otimes \bm{1}_M,\bm{v} \otimes \bm{e_1})\le M\cdot w_H(\bm{u},\bm{v}) & = & Md. & \IEEEQEDhere
    \end{IEEEeqnarray*}
    
	\end{IEEEproof}
	
	 The next example also demonstrates that the dual-rail construction may sometimes increase minimum distance, and this may be a reason to investigate $M > 2$ in the above construction, where the distance $d'$ satisfies $d\le d'\le Md$ (Theorem \ref{distance}).
	 
	 
	 	\begin{example}
		\label{example: 5,1,3}
		\normalfont
		Consider the $[[5,1,3]]$ stabilizer code with generator matrix $G_{\mathcal{S}}=[A|B]$ where
		\begin{align}
		A = 
		\left[
		\begin{array}{ccccc}
		1 & 0 & 0 & 1 & 0 \\
		0 & 1 & 0 & 0 & 1 \\
		1 & 0 & 1 & 0 & 0 \\
		0 & 1 & 0 & 1 & 0 \\
		\end{array}
		\right]
		\text{ and }
		B= 
		\left[
		\begin{array}{ccccc}
		0 & 1 & 1 & 0 & 0  \\
		0 & 0 & 1 & 1 & 0  \\
		0 & 0 & 0 & 1 & 1  \\
		1 & 0 & 0 & 0 & 1  \\
		\end{array}
		\right].
		\end{align}
		The code is not a CSS code. The stabilizer code derived from our construction has generator matrix 
		\begin{align}
		\setlength\aboverulesep{0pt}\setlength\belowrulesep{0pt}
		\setlength\cmidrulewidth{0.5pt}
		G_{\mathcal{S}'} = 
		\begin{blockarray}{ccc}
		& & \text{signs}\\ 
		\begin{block}{[c|c]c}
		A\otimes [1,1] & B\otimes [1,0]  & +\\ 
		\cmidrule(lr){1-2}
		 & I_5 \otimes [1,1] & -\\ 
		\end{block} 
		\end{blockarray}.
		\end{align}
		Consider $(\bm{y},\bm{x})$ such that $(\bm{x},\bm{y})$ is not in the row space of $G_{\mathcal{S}'}$ and $G_{\mathcal{S}'}(\bm{y},\bm{x})^T = 0$. We observe that $2\mid w_H(x)$. If $\bm{x}=\bm{0}$, then $\bm{y} = \bm{w}\otimes [1,1] \oplus \bm{1}_5 \otimes [1,0]$ for some $\bm{w}\in\F_2^5$, then after possibly applying the cyclic symmetry, we may assume $\bm{x}=\bm{e_1} \oplus \bm{e_2}$ and $(A\otimes [1,1])\bm{y}^T = [0,0,0,1]^T$. We observe that neither $[0,0,0,1]$ nor $[1,0,1,0]\oplus[0,0,0,1]=[1,0,1,1]$ is a column of $A$. It follows that the distance $d'\ge 4$. In fact, we see $d'=4$ by taking 
		\begin{equation}
		(\bm{x}',\bm{y}')=[1, 1, 0, 0, 0, 0, 0, 0, 0, 0|0, 0, 1, 0, 0, 0, 0, 0, 1, 0].
		\end{equation}
		Hence, the stabilizer code derived from the above construction has parameters $[[10,1,4]]$.
		
		By choosing $y$ to be either $[0,1]$ or $[1,0]$ for each of the five connected components with size $M=2$, we ensure $\mathcal{V}(\mathcal{S}')$ to satisfy Theorem~\ref{thm:trans_Z_all_l}, and thus it is oblivious to coherent noise.  
	\end{example}
    We now consider the cases that when some qubits are not involved in any $X$-stabilizer.
    
    \begin{example}
    \label{exam:5_1_2}
    \normalfont
        Consider the $[[5,1,2]]$ CSS code with the character vector $\bm{y} = [1,0,1,0,1]$ defined by the following generator matrix
        \begin{align}
		\setlength\aboverulesep{0pt}\setlength\belowrulesep{0pt}
		\setlength\cmidrulewidth{0.5pt}
		G_S = 
		\left[
		\begin{array}{ccccc|ccccc}
		1 & 1 & 1 & 1 & 0 &   &   &   &   &    \\
		\hline 
		  &   &   &   &   & 1 & 1 & 0 & 0 & 0  \\
		  &   &   &   &   & 0 & 0 & 1 & 1 & 0  \\
		  &   &   &   &   & 0 & 0 & 0 & 0 & 1  \\
		\end{array}
		\right].
		\end{align}
		Here, we have two connected components $\Gamma_1 = \{1,2\}$ and $\Gamma_2 = \{3,4\}$. Since $\mathrm{supp}([1,1,1,1,0]) = \Gamma_1 \cup \Gamma_2$, and $w_H(\bm{y_k}) = \frac{|\Gamma_k|}{2}=1$ for $k=1,2$, the two conditions in Theorem \ref{thm:trans_Z_all_l} are satisfied. Hence, the $[[5,1,2]]$ CSS code is oblivious to coherent noise, and we use \eqref{eqn:gen_encode_map} to compute computational states to verify it is a constant excitation code:
 		\begin{equation}
 		    \ket{\bar{0}} = \frac{1}{\sqrt{2}} (\ket{01011}+ \ket{10101}),
 		\end{equation}
 		\begin{equation}
 		    \ket{\bar{1}} = \frac{1}{\sqrt{2}} (\ket{10011}+ \ket{10101}).
 		\end{equation}
 		Here, the constant excitation is $3\neq \frac{5}{2}$ (half of the number of physical qubits). After the concatenation, we may introduce extra physical qubits by adding zeros to the current $X$-stabilizers and including all weight $1$ $Z$-stabilizers on the extra qubits. This construction reduces rate, but provides a large class of codes that may be useful in implementing logical gates. 
    \end{example}
	
	Given any $[[n,k,d]]$ stabilizer code, the theoretical construction in \eqref{three_part} and the observation in Example \ref{exam:5_1_2} provide a $[[Mn+s,k,d']]$ stabilizer code that is oblivious to coherent noise, where $d\le d'\le Md$, $M\ge 2$ is even, and $s \ge 0$.

	\section{Conclusion}
	\label{sec:conclusion}
	We derived necessary and sufficient conditions for a stabilizer to be oblivious to coherent noise, We showed that a CSS code that is oblivious to coherent noise must be a constant excitation code. These results were obtained by analyzing stabilizer codes for which the code space is preserved by transversal $\pi/2^l$ $Z$-rotations for all $l\ge2$. We intend to investigate the finite length setting, where the code space is only preserved by transversal $\pi/2^l$ $Z$-rotations for $l\le l_{\max}$. We expect these codes to prove useful in fault-tolerant implementations of non-Clifford gates. 

    \appendices
    \numberwithin{equation}{section}
    \renewcommand{\theequation}{A\arabic{equation}}
	\section{Proofs for Some Results}

	\subsection{Proof for Logical Identity induced by infinite transversal $Z$-rotations}
	\label{sec:proof_logical_identity}

	Assume $S$ defines an error-detecting code $[[n,n-r,d]]$, i.e., $d\ge 2$, which is invariant under all the transversal $\frac{\pi}{2^l}$ $Z$-rotations. Set $\theta_l = \frac{\pi}{2^l}$. Then, we can write the Taylor expansion 
	\begin{align}
	&\bigotimes_{i=1}^{n} e^{\imath\theta_l Z_i} 
	 = \bigotimes_{i=1}^n \sum_{k=0}^{\infty} \frac{(\imath\theta_l Z_i)^k}{k!}  
	= \bigotimes_{i=1}^n(I_2 + \imath\theta_lZ_i + \mathcal{O}(\theta_l^2)I_2) \\
	& = I_{2^n} + \imath\theta_l(Z_1 \otimes I_2 \otimes \cdots I_2 + I_2 \otimes Z_2 \otimes I_2 \otimes \cdots \otimes I_2 \nonumber\\
	&~~~~~~~~~~~~~~~~ + \cdots + I_2 \otimes I_2 \otimes \cdots \otimes Z_n) + \mathcal{O}(\theta_l^2)I_{2^n}.
	\end{align}
	We can choose $l$ large enough (say $l\ge L$ for some positive integer $L$) in order to ignore the last term,
	\begin{align}
	& \bigotimes_{i=1}^{n} e^{\imath\theta_lZ_i} \nonumber \\
	&\approx I_{2^n} + \imath\theta_l(Z_1 \otimes I_2 \otimes \cdots I_2 + I_2 \otimes Z_2 \otimes I_2 \otimes \cdots \otimes I_2 \nonumber \\
	&~~~~~~~~~~~~~~~~ + \cdots + I_2 \otimes I_2 \otimes \cdots \otimes Z_n).
	\end{align}
	On one hand, since the code can detect any single-qubit error, it can detect any linear combination of them (Theorem 10.2 in \cite{nielsen2011quantum}). Therefore, $\bigotimes_{i=1}^{n} e^{\imath\theta_lZ_i} $ is detectable (i.e., it maps all the codewords outside the codespace or acts trivially on the codespace). On the other hand, $\bigotimes_{i=1}^{n} e^{\imath\theta_lZ_i} $ preserves the code space by assumption. Therefore,  $\bigotimes_{i=1}^{n} e^{\imath\theta_lZ_i} $ act trivally on the codespace, which implies that the logical operator induced by $\bigotimes_{i=1}^{n} e^{\imath\theta_lZ_i} $ is identity for all $l\ge L$. Note that the logical operator induced by $\bigotimes_{i=1}^{n} e^{\imath\theta_lZ_i} $ is identity for larger $l$ implies that the logical operator induced by $\bigotimes_{i=1}^{n} e^{\imath\theta_lZ_i} $ is also identity for smaller $l$ via repeated applications. Therefore, the logical operator induced by $\bigotimes_{i=1}^{n} e^{\imath\theta_lZ_i} $ is identity for all $l$.
	\hfill \IEEEQEDhere

 \subsection{Proof of Lemma \ref{halfonesomeodd}}
  \renewcommand{\theequation}{B\arabic{equation}}
    \label{sec:proof_of_odd_wt}
        We may assume that $\bm{y} \neq \bm{0},\bm{1}$, and that the subspaces $\mathcal{W},\bm{y}^\perp$ and their duals $\langle \bm{1} \rangle $, $\langle \bm{y} \rangle $ intersect as shown below. The edge label is the index of the smaller subspace in the group larger subspace.
		\begin{center}
			\begin{tikzpicture}
			\node (A) at (1.5,5) {$\F_2^{m}$};
			\node (B) at (0,3) {$\mathcal{W}$};
			\node (C) at (3,3) {$\bm{y}^\perp$};
			\node (D) at (1.5,1) {$\mathcal{W} \cap \bm{y}^\perp$};
			
			\path  (A) edge node[left] {$2$} (B);
			\path  (B) edge node[left] {$2$} (D);
			\path  (A) edge node[right] {$2$} (C);
			\path  (C) edge node[right] {$2$} (D);
			
			\node (E) at (6.5,5) {$\langle \bm{1},\bm{y}\rangle$};
			\node (F) at (5,3) {$\langle \bm{1}\rangle$};
			\node (G) at (8,3) {$\langle \bm{y} \rangle$};
			\node (H) at (6.5,1) {$\langle \bm{0} \rangle$};
			
			\path  (E) edge node[left] {$2$} (F);
			\path  (F) edge node[left] {$2$} (H);
			\path  (E) edge node[right] {$2$} (G);
			\path  (G) edge node[right] {$2$} (H);
			
			\end{tikzpicture}
		\end{center}
		We have
		\begin{align}\label{F_minus_W}
		&\frac{\sum_{\bm{v}\in\F^m_2 \setminus \mathcal{W}} \epsilon_{\bm{v}} \left(\imath\tan\frac{2\pi}{2^l}\right)^{w_H(\bm{v})}}{\left(\sec\frac{2\pi}{2^l}\right)^m} \nonumber\\
		&= P\left[\left(\F^{m}_2 \setminus \mathcal{W}\right) \cap \bm{y}^\perp\right] - P\left[ \left(\F^{m}_2 \setminus \mathcal{W}\right)\cap \left(\F^{m}_2 \setminus \bm{y}^\perp\right)\right].
		\end{align}
		Table \ref{part_F_minus_W} specifies how subsets $T$ appearing \eqref{F_minus_W} can be expressed as disjoint unions of subsets $A$ that appear in the MacWilliams Identities.
		\begin{table*}
			\begin{center}
			\caption{ Sign patterns for different weight enumerators $P[A]$ with $A \subset \F^m_2$: the entries of each row specify how the set corresponding to the subsets $A$ can be written as a union of subsets in different columns.}
			\label{part_F_minus_W}
			\begin{tabular}{|c |c |c |c |} 
				\hline
				\diagbox[width=7.5em]{$A$}{$T$}& $ (\F^{m}_2 \setminus \mathcal{W})\cap (\F^{m}_2 \setminus \bm{y}^\perp)$  & $(\F^{m}_2 \setminus \mathcal{W}) \cap \bm{y}^\perp$& $\mathcal{W} \cap (\F^{m}_2 \setminus \bm{y}^\perp)$ \\
				\hline
				$\F^{m}_2 \setminus \mathcal{W}$ & $+$ & $+$ & $0$ \\ 
				\hline
				$\F^{m}_2 \setminus \bm{y}^\perp$ & $+$ & $0$ & $+$ \\
				\hline
				$W \setminus (\mathcal{W} \cap \bm{y}^\perp)$ & $0$ & $0$ & $+$ \\ 
				\hline
			\end{tabular}
			\end{center}
		\end{table*}
		It follows from Table \ref{part_F_minus_W} that we may rewrite the right hand side of (\ref{F_minus_W}) as
		\begin{align}\label{partition}
		&\frac{\sum_{\bm{v}\in\F^m_2 \setminus \mathcal{W}} \epsilon_{\bm{v}} \left(\imath\tan\frac{2\pi}{2^l}\right)^{w_H(\bm{v})}}{\left(\sec\frac{2\pi}{2^l}\right)^m} \nonumber \\ 
		&= P\left[\F^m_2 \setminus \mathcal{W}\right] - 2P\left[\F^m_2\setminus \bm{y}^\perp\right] + 2P\left[\mathcal{W}\setminus (\mathcal{W}\cap \bm{y}^\perp)\right].
		\end{align}
		It follows from (\ref{P_F_minus_W}) that
		\begin{equation} \label{first_term}
		P\left[\F^m_2 \setminus \mathcal{W}\right] = \imath\sin\frac{2\pi m}{2^l}.
		\end{equation}
		We rewrite (\ref{P_F}) as
		\begin{equation}
		P\left[\F^m_2 \setminus \bm{y}^\perp\right] = e^{\imath\frac{2\pi m}{2^l}} - P[y^\perp].
		\end{equation}
		Recall that we define $t_{+} = \cos\frac{2\pi}{2^l}+\imath\sin{\frac{2\pi}{2^l}}$ and $t_{-}= \cos\frac{2\pi}{2^l}-\imath\sin{\frac{2\pi}{2^l}}$. We apply the MacWilliams Identities to obtain 
		\begin{align}
		P\left[y^\perp\right] 
		& = \frac{1}{|\langle \bm{y} \rangle|}P_{|\langle \bm{y} \rangle|}\left(t_{+},t_{-}\right) \nonumber \\
		& = \frac{1}{2}\left(e^{\imath\frac{2\pi m}{2^l}} + e^{\imath\frac{2\pi(m-2w_H(\bm{y}))}{2^l}}\right),
		\end{align}
		so that 
		\begin{equation} \label{second_term}
		P\left[\F^m_2 \setminus \bm{y}^\perp \right] = \frac{1}{2} \left(e^{\imath\frac{2\pi m}{2^l}} - e^{\imath\frac{2\pi(m-2w_H(\bm{y}))}{2^l}}\right).
		\end{equation}
		It follows from (\ref{P_W}) that 
		\begin{equation}
		P\left[\mathcal{W}\setminus (\mathcal{W}\cap \bm{y}^\perp)\right] = \cos\frac{2\pi m}{2^l} - P[\mathcal{W}\cap \bm{y}^\perp].
		\end{equation}
		We apply the MacWilliams Identities to obtain 
		\begin{align}
		& P\left[\mathcal{W}\cap \bm{y}^\perp\right] \nonumber\\
		& = \frac{1}{|\langle \bm{1},\bm{y} \rangle|}P_{|\langle \bm{1},\bm{y} \rangle|}\left(t_{+},t_{-}\right)\nonumber \\
		& = \frac{1}{4}\left[ e^{\imath\frac{2\pi m}{2^l}} + e^{-\imath\frac{2\pi m}{2^l}} + e^{\imath\frac{2\pi (m-2w_H(\bm{y}))}{2^l}} + e^{\imath\frac{2\pi (2w_H(\bm{y})-m)}{2^l}} \right] 
		\end{align}
		so that 
		\begin{equation}\label{third_term}
		P\left[\mathcal{W}\setminus(\mathcal{W}\cap \bm{y}^\perp)\right] =\frac{1}{2}\left[\cos\frac{2\pi m}{2^l}  - \cos\frac{2\pi (m-2w_H(\bm{y}))}{2^l} \right].
		\end{equation}
		We now use (\ref{first_term}), (\ref{second_term}), (\ref{third_term}) to rewrite the right hand side of (\ref{partition}) as
		\begin{align}
		&\imath\sin\frac{2\pi m}{2^l} - e^{\imath\frac{2\pi m}{2^l}} + e^{\imath\frac{2\pi(m-2w_H(\bm{y}))}{2^l}} + \cos\frac{2\pi m}{2^l} \nonumber\\
		&- \cos\frac{2\pi (m-2w_H(\bm{y}))}{2^l},
		\end{align}
		which reduces to \eqref{halfonesomeodd_conclusion}.
    \hfill \IEEEQEDhere

    \section*{Acknowledgment}
	We would like to thank Kenneth Brown, Dripto Debroy, and Michael Newman for helpful discussions. 
	We would like to thank our Associate Editor Salman Beigi and the two reviewers for providing valuable feedback, which improved the presentation of our results significantly. 

	\bibliographystyle{IEEEtran}


\begin{IEEEbiographynophoto}{Jingzhen Hu}
(Graduate Student Member, IEEE) received the B.S. degree with honors in mathematics from Southern Methodist University, USA, in 2017. She is currently pursuing the Ph.D. degree in mathematics at Duke university. Her current research interests include developing mathematical tools from algebra and combinatorics to solve problems in quantum error correction, quantum information theory, and quantum algorithms.

She received the Hamilton undergraduate research scholar award, summer research award, and statistical science department award for academic excellence at Southern Methodist University in 2017. She became a member of Phi Beta Kappa, Gamma of Texas since 2017. She also received Carrie \& Edwin Mouzon Mathematics scholarship (2015-2017), Founders scholarship, Discovery scholarship, and Mustang scholar award (2014-2017) at Southern Methodist University. She is currently the V.P. in the Duke student chapter of SIAM. 
\end{IEEEbiographynophoto}

\begin{IEEEbiographynophoto}{Qingzhong Liang}
(Graduate Student Member, IEEE) received the B.S. degree (high honors with distinction) in mathematics from the University of Michigan - Ann Arbor, USA, in 2017. He is currently pursuing the Ph.D. degree in mathematics at Duke university. His current research interests include combinatorics, quantum error correction, and quantum computing. His personal website is \url{http://www-personal.umich.edu/~qzliang/}.

He ranked 15th in the nation and 1st in the department at the 37th Virginia Tech Regional Mathematics Contest. He also ranked 260.5th in the nation and 3rd in the department at the 75th William Lowell Putnam Mathematical Competition. He is currently an officer in the Duke student chapter of SIAM, organizing the 2021 Triangle Area Graduate Mathematics Conference (TAGMaC).
\end{IEEEbiographynophoto}

\begin{IEEEbiographynophoto}{Narayanan Rengaswamy}
(Member, IEEE) is a postdoctoral research associate with Prof. Bane Vasic at the University of Arizona, where he is involved in the error correction aspects of the NSF funded Center for Quantum Networks (CQN) and DoE funded Superconducting Quantum Materials and Systems (SQMS) center. He completed his Ph.D. in Electrical Engineering in May 2020 at Duke University, working under the supervision of Prof. Henry Pfister and Prof. Robert Calderbank. His dissertation (https://arxiv.org/abs/2004.06834) focused on developing systematic methods to construct fault-tolerant logical operations on stabilizer quantum error correcting codes, and on optimally decoding classical codes over the quantum pure-state channel that arises in free-space optical communications. Prior to this, he completed his M.S. in Electrical Engineering in December 2015 at Texas A\&M University, where he worked with Prof. Henry Pfister on cyclic polar codes. In summer 2015, he was a research intern at Alcatel-Lucent Bell Labs, Stuttgart, Germany, where he analyzed the finite-length performance of spatially-coupled LDPC codes on the binary erasure channel, under the supervision of Dr. Laurent Schmalen and Dr. Vahid Aref. His general research interests are in classical and quantum information theory, coding theory, compressed sensing and statistical inference problems. He is passionate about discovering connections between the classical and quantum information processing worlds.
\end{IEEEbiographynophoto}

\begin{IEEEbiographynophoto}{Robert Calderbank}
(Fellow, IEEE) received the B.S. degree from Warwick University, U.K., in 1975, the M.S. degree from Oxford University, U.K., in 1976, and the Ph.D. degree from the California Institute of Technology, in 1980, all in mathematics.

He is currently a Professor of electrical and computer engineering with Duke University, where he directs the Rhodes Information Initiative at Duke (iiD). Prior to joining Duke in 2010, he was a Professor of electrical engineering and mathematics with Princeton University. Prior to joining Princeton in 2004, he was the Vice President for Research at AT\&T, responsible for directing the first Industrial Research Laboratory in the world where the primary focus is data at scale. At the start of his career at Bell Labs, innovations by Dr. Calderbank were incorporated in a progression of voiceband modem standards that moved communications practice close to the Shannon limit. Together with Peter Shor and colleagues at AT\&T Labs, he developed the mathematical framework for quantum error correction. He is a co-inventor of space-time codes for wireless communication, where correlation of signals across different transmit antennas is the key to reliable transmission. He was a member of the Board of Governors of the IEEE Information Theory Society from 1991 to 1996 and from 2006 to 2008. He was honored by the IEEE Information Theory Prize Paper Award in 1995 for his work on the Z4 linearity of Kerdock and Preparata Codes (joint with A.R. Hammons Jr., P.V. Kumar, N.J.A. Sloane, and P. Sole), and again in 1999 for the invention of space-time codes (joint with V. Tarokh and N. Seshadri). He has received the 2006 IEEE Donald G. Fink Prize Paper Award, the IEEE Millennium Medal, the 2013 IEEE Richard W. Hamming Medal, and the 2015 Shannon Award. He was elected to the U.S. National Academy of Engineering in 2005. He has served as the Editor-in-Chief for the IEEE TRANSACTIONS ON INFORMATION THEORY from 1995 to 1998 and an Associate Editor for Coding Techniques from 1986 to 1989.

\end{IEEEbiographynophoto}

\end{document}